\documentclass[aps,twocolumn,amssymb,showpacs]{revtex4}
\usepackage{graphicx}
\usepackage{multirow}
\usepackage[breaklinks,colorlinks,urlcolor=blue,citecolor=blue]{hyperref}
\usepackage{mathrsfs}
\usepackage{amsmath}
\usepackage{color}

\begin{document}

\title{Constraining Superfluidity in Dense Matter from
the Cooling of\\ Isolated Neutron Stars} 
\author{Spencer Beloin$^{1}$}
\author{Sophia Han$^{1}$} 
\author{Andrew W. Steiner$^{1,2}$} 
\author{Dany Page$^{3}$}
\affiliation{$^{1}$Department of Physics and Astronomy, University of
  Tennessee, Knoxville, TN 37996, USA}
\affiliation{$^{2}$Physics Division, Oak Ridge National Laboratory, Oak
  Ridge, TN 37831, USA}
\affiliation{$^{3}$Instituto de Astronom\'{i}a, Universidad Nacional
  Aut\'{o}noma de M\'{e}xico, Mexico D.F. 04510, Mexico}

\begin{abstract}
  We present a quantitative analysis of superfluidity and
  superconductivity in dense matter from observations of isolated
  neutron stars in the context of the minimal cooling model. Our new
  approach produces the best fit neutron triplet superfluid critical
  temperature, the best fit proton singlet superconducting critical
  temperature, and their associated statistical uncertainties. We find
  that the neutron triplet critical temperature is likely
  $2.09^{+4.37}_{-1.41} \times 10^{8}$ K and that the proton singlet
  critical temperature is $7.59^{+2.48}_{-5.81} \times 10^{9}$ K.
  However, we also show that this result only holds if the Vela
  neutron star is not included in the data set. If Vela is included,
  the gaps increase significantly to attempt to reproduce
  Vela's lower temperature given its young age. Further including
  neutron stars believed to have carbon atmospheres increases the
  neutron critical temperature and decreases the proton critical
  temperature. Our method demonstrates that continued observations of
  isolated neutron stars can quantitatively constrain the nature of
  superfluidity in dense matter.
\end{abstract}


\maketitle

\section{Introduction}

Neutron stars, the remnants of the gravitational collapse of $\sim{8}$
to $20~\mathrm{M}_{\odot}$ main-sequence stars, contain matter with
densities at least several times larger than the densities at the
center of atomic nuclei~\cite{Lattimer01}. Matter at these densities
is difficult to probe in the laboratory, except at high temperatures,
which confounds the extraction of dense matter properties from
experiment. Thus, neutron stars are a unique laboratory for the study
of dense and strongly-interacting matter.

Current constraints from neutron star mass and radius observations
determine the equations of state of dense matter (EOS) above the
nuclear saturation densities to within about a factor of two (see
recent constraints in Refs.~\cite{Steiner15un,Nattila16eo} or an
alternate perspective in Ref.~\cite{Ozel16}). Recent progress in
nuclear theory constrains the energy per baryon of neutron matter at
the saturation density to within a few MeV~\cite{Gandolfi15nm}.
However, the EOS alone is not enough to fully describe dense matter.
Almost all neutron star observables also require some knowledge of how
energy and momentum are transported in dense matter. Transport
properties, in turn, are strongly affected by the presence of
superconductivity and superfluidity~\cite{Page14ss}.

At the end of a supernova, the neutron star is born with a core
temperature $\sim$ $10^{11}$ K, and, in some cases, a measurable
velocity with respect to the remnant. Except for a thin shell at the
surface, the neutron star becomes isothermal after a few hundred
years. In isolated neutron stars without a companion, the temperature
decreases (unless heated by magnetic field dissipation or some dark
matter-related process) at a rate determined by the nature of dense
matter~\cite{Page04,Yakovlev04,Page06}. In the first $10^5$ years,
cooling is dominated by the emission of neutrinos from the core, after
which photon emission from the surface takes over. The neutrino rates
strongly depend on the nature of neutron superfluidity and proton
superconductivity. Thus, if one obtains temperature and age estimates
from a number of cooling neutron stars, the comparison of theoretical
models to data results in a constraint on the nature of superfluidity
in dense matter.

\section{Method}

There are several isolated neutron stars where age estimates are
available and where x-ray data provides an estimate of the surface
temperature. The extraction of the surface temperature, however,
depends on the composition of the atmosphere. Older neutron stars are
expected to have atmospheres made of iron-peak elements and these
atmospheres are well fit by black body models giving black body radii
in the range of 10$-$13 km expected from theoretical
models~\cite{Steiner16ns}. The inferred radii from black body fits to
younger stars are often much smaller than expected, leading to the
idea that younger isolated neutron stars may have light-element
atmospheres, and hydrogen (H) atmosphere fits to the data often result in neutron
star radii closer to what is expected. For most objects, only black
body and H atmosphere fits to the x-ray data are available.

The temperature profile of the star depends on the composition of the
envelope, which is the region between the photosphere and a boundary
density near $\rho_b=10^{10}~\mathrm{g}/\mathrm{cm}^3$. This boundary density
is defined so that the luminosity at this boundary is equal to the
total luminosity of the star. In the case of a light-element
atmosphere, the presence of light elements in the envelope can modify
the inferred surface temperature. Light-element envelopes are not
expected with iron-peak atmospheres described by black body models, as
light elements in the envelope will inevitably make their way to the
surface.

Similar to the procedure used in Ref.~\cite{Page04}, we use the
temperatures and luminosities implied by H atmosphere fits to the
x-ray spectra for younger stars (less than about $10^5$ years) in
which black body radii are too small to be realistic. In older stars,
we use temperatures and luminosities obtained by blackbody fits to the
x-ray spectra. The observational data set is summarized in
Table~\ref{tab:data}. In the case of PSR J2043, we use the results
from an H atmosphere fit because no blackbody fit was available.

The true x-ray spectrum of an isolated neutron star is not that of a
black body. Modeling heat transport in hydrostatic equilibrium,
Ref.~\cite{Gudmundsson83} found one can obtain a simple relationship
between the effective surface temperature which depends on the amount
of light elements in the envelope and the temperature at the base of
the envelope, $T_b=T(\rho_b)$. This relationship simplifies the
calculation considerably, allowing one to connect envelope models on
top of a neutron star interior~\cite{Page04}. Younger stars
may have light elements which affects the surface temperature, but
older stars which have heavy element photospheres are not expected to
have light element envelopes (as otherwise the light elements in the
envelope would move towards the surface). The work in
Ref.~\cite{Gudmundsson83} was updated in Ref.~\cite{Potekhin97}
and our neutron star cooling model uses this work to determine the
effective surface temperature and luminosity as a function of the
temperature at the base of the envelope. We vary the amount of light
elements in the envelope in all neutron stars less than $10^5$ years
old, which is consistent with the notion that neutron star atmospheres
evolve from light elements to iron-peak elements over time through
nuclear fusion.

We also do not include any stars with magnetic fields larger than
$10^{14}$ G in our data set, as the magnetic field has a strong impact
on the atmosphere and may cause strong variations of the temperature
on the surface which our model cannot accurately
describe~\cite{Potekhin:2014hja}. In some cases, H atmosphere
fits to x-ray spectra imply magnetic fields on the order of $10^{12}$
G, but we assume that there is no modification to the surface
temperature or luminosity from these fields. In particular, we
assume the temperature distribution is uniform across the neutron star
surface.

There are a few objects for which neither H nor black body atmospheres
imply a realistic neutron star radius, but where carbon atmospheres
fit well. This is the case for the neutron star located in Cassiopeia
A and XMMU J1732 located in HESS J1731$-$347, which we include in our
analysis along with the possibility that they also may contain light
elements in their envelopes.

If a neutron star can be associated with a nearby supernova remnant
and its proper motion can be measured, one can determine the kinetic
age, $t_{\mathrm{kin}}$. Alternatively, pulsar ages can be estimated
from the spin-down timescale, $t_{\mathrm{sd}}=P/(2\dot{P})$, an age
estimate assuming an evolution with a dipolar magnetic field.
Spin-down ages can be measured precisely, but they disagree with
kinetic ages by a factor of 3 or more~\cite{Page09ne}, thus we assume
a factor of 3 uncertainty in $t_{\mathrm{sd}}$. We presume kinetic
ages are more accurate than spin-down ages, but this is not certain.

\begin{table*}
\begin{tabular}{lcccccccc}
\multirow{2}{*}{Star} & $\log_{10} (t_{\rm{sd}}/\mathrm{yr})$ &
\multirow{2}{*}{$B$ (G)} &
\multirow{2}{*}{$\log_{10}(T^{\infty}/\mathrm{K})$} &
atmos. & $\log_{10}(L^{\infty})$ & mass &
radius & \multirow{2}{*}{Ref.} \\
& or $\log_{10} (t_{\rm{kin}}/\mathrm{yr})$ & & &
model & $\mathrm{erg}/\mathrm{s}$ & ($M_{\odot}$) & (km) & \\
\hline

Cas A NS & 2.52 (observed) \cite{Fesen:2006zma} & $8 \times 10^{10}$ &
$6.26^{+0.02}_{-0.02}$ & C & $33.63^{+0.08}_{-0.08}$ & $1.4^{*}$ & 12-15&
\cite{Ho09} \\

& & &
$6.447^{+0.012}_{-0.012}$ & H &  & $1.4^{*}$ & 4&
\cite{Ho09} \\

& & &
$6.653^{+0.007}_{-0.007}$ & BB &  & $1.4^{*}$ & $<$ 1&
\cite{Ho09} \\

PSR J1119$-$6127 & 3.20 (s) \cite{Kumar2012} & $4.1\times 10^{13}$ &
$6.09^{+0.08}_{-0.08}$ & HA & $33.32^{+0.14}_{-0.14}$ & $1.4^{*}$ & 10$^{*}$ &
\cite{SafiHarb08} \\

& & &
$6.39^{+0.02}_{-0.02}$ & BB & $33.39^{+0.11}_{-0.11}$ &  $1.4^{*}$ & $2.7{\pm}0.7$
& \cite{SafiHarb08} \\

RX J0822$-$4247 & $3.57^{+0.04}_{-0.04}$ (k)  &
$\sim10^{12}$ &
$6.24^{+0.04}_{-0.04}$ & HA & $33.93^{+0.08}_{-0.08}$ &  $1.4^{*}$ & 10$^{*}$ &
\cite{Zavlin:1999} \\ 

& & &
$6.65^{+0.04}_{-0.04}$ & BB & $33.75^{+0.15}_{-0.15}$ &$1.4^{*}$ & $\approx$2&
\cite{Zavlin:1999} \\ 

1E 1207.4$-$5209$^{++}$ & $3.85^{+0.48}_{-0.48}$ (k)  \cite{Zavlin00,Roger98} &
$3\times 10^{12}$  &  
$6.21^{+0.07}_{-0.07}$ & HA & $33.50^{+0.24}_{-0.24}$ & $1.4^{*}$ & 10$^{*}$ &
\cite{Pavlov:2002b} \\
& & &
$6.48^{+0.01}_{-0.01}$ & BB & $33.29^{+0.59}_{-0.59}$ & $1.4^{*}$ & $<$ 1.5&
\cite{Mereghetti96,Zavlin98} \\

PSR J1357$-$6429 & 3.86 (s) & $8{\times}10^{12}$ & 
$5.88^{+0.04}_{-0.04}$ & HA & $32.63^{+0.17}_{-0.17}$ & $1.5-1.6$ & 10$^{*}$ &
\cite{Zavlin:2007nx} \\ 
& & &
$6.23^{+0.05}_{-0.05}$ & BB & $33.56^{+0.20}_{-0.20}$  & $1.5-1.6$
& $2.5{\pm}0.5$ &
\cite{Zavlin:2007nx} \\ 

RX J0002+6246$^{\dagger}$ & $3.96^{+0.08}_{-0.08}$ (k) & &
$6.03^{+0.03}_{-0.03}$ & HA & $33.21^{+0.13}_{-0.13}$ & & &
\cite{Page04,Pavlov04} \\ 
& & &
$6.15^{+0.11}_{-0.11}$ & BB & $32.5^{+0.32}_{-0.32}$ & & &
\cite{Page04,Pavlov04} \\

PSR B0833$-$45 & $3.97^{+0.23}_{-0.23}$ 
(k)~\cite{Tsuruta09} & $3{\times}10^{12}$ & 
$5.83^{+0.02}_{-0.02}$ & HA & $32.58^{+0.04}_{-0.04}$ & $1.4^{*}$ & 13&
\cite{Pavlov:2001hp} \\
& & &
$6.18^{+0.02}_{-0.02}$ & BB & $32.16^{+0.12}_{-0.12}$ & $1.4^{*}$ & $2.1{\pm}0.2$&
\cite{Pavlov:2001hp} \\

PSR B1706$-$44 & 4.24 (s) \cite{Gotthelf02} & $3{\times}10^{12}$ &
$5.80^{+0.13}_{-0.13}$ & HA & $32.37^{+0.56}_{-0.56}$ & $1.45-1.59$ & 13&
\cite{McGowan04} \\
& & & $6.22^{+0.04}_{-0.04}$ & BB & $32.78^{+0.30}_{-0.30}$ & $1.4^{*}$ &
$<6$& \cite{Gotthelf02} \\

XMMU J1732$-$344 & $4.43^{+0.17}_{-0.43}$ (k)~\cite{Tian:2008tr} &
$\sim 10^{10-11}$ & $6.25^{+0.01}_{-0.0045}$ & C &
$33.99^{+0.04}_{-0.02}$ & & & \cite{Klochkov:2014ola} \\ 

PSR J0538+2817$^{\|}$ & $4.47^{+0.05}_{-0.06}$
(s)~\cite{Kramer:2003au} & $\sim 10^{12}$ &
$6.05^{+0.10}_{-0.10}$ & HA & $33.10^{+0.50}_{-0.50}$ & $1.4^{*}$
& 10.5& \cite{Zavlin04c} \\ 
& & & $6.327^{+0.007}_{-0.007}$ & BB & & $1.4^{*}$ &
$<$ 2& \cite{McGowan03} \\ 

PSR B2334+61 & 4.61 (s) & $\sim10^{10-12}$ &
$5.68^{+0.17}_{-0.17}$ & HA & $32.70^{+0.68}_{-0.68}$ & $1.4^{*}$ & 10-13
& \cite{McGowan06} \\
 & & &
$6.02^{+0.19}_{-0.19}$ & BB & & $1.4^{*}$ & $<$ 2
& \cite{McGowan06} \\

PSR B0656+14 & 5.04 (s)~\cite{Zavlin:2007nc} &
$5{\times}10^{12}$ \cite{Mignani15} &
$5.71^{+0.03}_{-0.04}$ & BB & $32.58^{+0.40}_{-0.40}$  & $1.4^{*}$ & 12-17
& \cite{Possenti:1996em} \\ 

PSR J1740+1000 & 5.06 (s) \cite{McLaughlin:2001kh} &
$1.8{\times}10^{12}$ & $6.04^{+0.01}_{-0.01}$ 
& BB & $32.15^{+0.05}_{-0.05}$ & & & \cite{Kargaltsev:2012yi} \cite{Vigano13} \\

PSR B0633+1748 & 5.53 (s) &  &
$5.75^{+0.04}_{-0.05}$ & BB & $31.18^{+0.33}_{-0.33}$ & $1.4^{*}$ &
10$^{*}$ & \cite{Halpern1997a} \\ 

RX J1856.4$-$3754$^{\ddagger}$ &
$5.70^{+0.05}_{-0.25}$ (k) \cite{Ho:2007gs} &
$4{\times}10^{12}$ & $5.75^{+0.15}_{-0.15}$ & BB &
$31.56^{+0.12}_{-0.12}$ &  & 14 &
\cite{Pons02,Burwitz03} \\

PSR B1055$-$52$^{\mathsection}$ & 5.73 (s)  &
4${\times}10^{12}$ &  $5.88^{+0.08}_{-0.08}$ & BB &
$32.57^{+0.52}_{-0.52}$ & $1.4^{*}$ & 13&
\cite{Pavlov:2003da} \\ 

PSR J2043+2740$^{\mathparagraph}$ & 6.08 (s) & $4 \times 10^{11}$ & 
$5.64^{+0.08}_{-0.08}$ & HA & $29.62^{+0.52}_{-0.52}$ &  & 10$^{*}$ &
\cite{Zavlin:2007nc} \\ 

PSR J0720.4-3125 & $6.11$ (s)~\cite{deVries:2004jf} &
$~10^{13}$~\cite{Kaplan02b} &
$5.75^{+0.20}_{-0.20}$ & BB & $31.89^{+0.52}_{-0.52}$ & $1.4^{*}$ & 11-13&
\cite{Kaplan:2003hj} \\ 


\end{tabular}
\caption{The data set used in the current work is adapted from the
  earlier work in Refs.~\cite{Page04,Vigano13} and \cite{Lim15}. As in
  Ref.~\cite{Page04} we favor kinetic ages over spin-down ages where
  possible. The letters `s' and `k' in column 2 denote characteristic spin down 
  age and kinetic age, respectively.  
  References are given in column 2 only where our ages
  differ from the values used in Ref.~\cite{Page04}. We use \ H
  atmosphere (HA) fits to stars less than $10^{5}$ years and blackbody
  (BB) fits for older stars. In some of the H atmosphere fits, a
  magnetic field was used (either as a fixed value or as a fit
  parameter), and this is indicated in the fourth column (mHA). Notes:
  $(*)$ This value was assumed not derived. $(\|)$ For the H
  atmosphere fit, we use the redshifted temperature from
  Ref.~\cite{Zavlin04c}, $10^{6.04}$, instead of the value reported
  as $10^{5.94}$ in Ref.~\cite{Lim15}. $(\ddagger)$ As in
  Ref.~\cite{Page04}, we use a range determined by the colder
  blackbody component from Ref.~\cite{Pons02} and the warmer blackbody
  component in Ref.~\cite{Burwitz03}. $(\mathsection)$
  We have used
  the updated information from Ref.~\cite{Pavlov:2003da} as in
  Ref.~\cite{Lim15} over the values in Ref.~\cite{Page04}.
  $(\mathparagraph)$ We use a H atmosphere fit for this source
  since a blackbody fit is not available. $(5)$ As in
  Ref.~\cite{Page04} we use a range determined by the cold and warm
  components from the blackbody model in Ref.~\cite{Kaplan:2003hj}.
  $(\dagger)$ Ref.~\cite{Esposito08} claims this is not a neutron
  star. $(++)$ As discussed in
  Ref.~\cite{Page04}, Ref.~\cite{Zavlin04a} suggests
  that this star may be accreting due to its spin-down behavior.
}
\label{tab:data}
\end{table*}

We employ the minimal cooling model from Ref.~\cite{Page04}, assuming
that the neutron star is made entirely of neutrons, protons, and
leptons, and that the direct Urca process does not occur. When the
direct Urca process does not operate, the neutron star cooling depends
only weakly on the neutron star mass and the bulk thermodynamics of
matter which is determined by the equation of state. If the direct
Urca process does occur, then the cooling curves would depend strongly
on the equation of state and individual neutron star masses. In this
work, we assume the Akmal-Pandharipande-Ravenhall
(APR)~\cite{Akmal98eo} equations of state and we also set the mass of
all isolated neutron stars to 1.4 $\mathrm{M}_{\odot}$ (we will find
below that the data does not require the direct Urca process, except
possibly in the case of the Vela pulsar). We assume no additional
cooling occurs due from presence of deconfined quarks, Bose
condensates, or exotic (i.e., heavy) hadrons. The simplification
provided by the minimal cooling model is important because it allows
us to decrease our parameter space which is already relatively large
(as described below). We also ignore any possible effects on the
cooling from rotation.

In the minimal model, the principal unknown quantities in dense matter
which impact neutron star cooling are the neutron superfluid and
proton superconducting gaps. Superfluidity and superconductivity
exponentially suppress the specific heat and modify the neutrino
emissivities in dense matter (for a review see Ref.~\cite{Page14ss}).
These effects begin when the temperature of the neutron star cools
below the critical temperature. In the original BCS theory of
superconductivity, the critical temperature and the value of the gap
at zero temperature are related by $\Delta$$(\text{T}=0)$$\simeq1.8
\,k_B T_c$. The BCS approximation to superconductivity does not
necessarily apply in the strongly-interacting nucleonic fluid, but we
retain the standard practice of assuming that the BCS relation is
approximately correct.

Neutron superfluidity in the singlet ($^{1}\mathrm{S}_0$) channel is
present in the neutron star crust, but the critical temperatures are
too large to be constrained by the data of neutron stars older than a
few hundred years. Proton singlet superconductivity in the outer core
and neutron triplet $^{3}$P$_{2}$ superfluidity in the inner core, on the
other hand, are the most important parameters in the minimal cooling
model and can be constrained by neutron stars with the ages found in
our data set. Superfluid gaps suppress heat capacity for temperatures
well below $T_{c}$ (but increase heat capacity at temperature just
below $T_{c}$). Superfluidity and superconductivity also allow a new
neutrino emission process induced by the formation of Cooper pairs.
This cooling process is included, along with the correction from
suppression in the vector
channel~\cite{Flowers76a,Leinson06ne,Leinson06vc,Steiner09sr}. We also
include the axial anomalous contribution to the pair-breaking
emissivity from Ref.~\cite{Leinson:2009nu}.

Theoretical calculations of the neutron and proton critical
temperatures in the neutron star core appear approximately as Gaussian
functions of the Fermi momentum~\cite{Page14ss}. Pairing is suppressed at low
densities as the interparticle spacing is increased, and also
suppressed at high densities as the repulsion between nucleons
quenches the attractive interaction. In this work, we assume that both
the proton singlet and neutron triplet critical temperatures can be
described by the Gaussian form
\begin{equation}
  T_{c}(k_{F}) = T_{c,\mathrm{peak}} \exp\left[
    \frac{\left(k_{F}-k_{F,\mathrm{peak}}\right)^2}
         {2\Delta k_{F}^2}\right]
  \label{eq:gapparm}
\end{equation}
with parameters $T_{c,\mathrm{peak}}$, $k_{F,\mathrm{peak}}$ and
$\Delta k_{F}$. 

To avoid overcounting models where the gaps vanish, we
constrain $k_{F,p,\mathrm{peak}}$ and $k_{F,n,\mathrm{peak}}$ to lie
between the crust-core transition (taken to be at $n_B =
0.09~\mathrm{fm}^{-3}$) and the central density of a
$1.4~\mathrm{M}_{\odot}$ neutron star (at $n_B =
0.545~\mathrm{fm}^{-3}$). This implies $0.481~\mathrm{fm}^{-3} <
k_{F,p,\mathrm{peak}} < 1.304~\mathrm{fm}^{-3}$ and
$1.418~\mathrm{fm}^{-3} < k_{F,n,\mathrm{peak}} <
2.300~\mathrm{fm}^{-3}$. To avoid overcounting models where the gap is
nearly independent of density, we also enforce $\Delta k_{F,p} <
1.304~\mathrm{fm}^{-3}$ and $\Delta k_{F,n} <
2.300~\mathrm{fm}^{-3}$. Finally, we
constrain our critical temperatures to be smaller than $10^{10}$ K
because the fit is insensitive to larger values.
 
In the case where one is fitting a model to data with small
uncertainties in the independent variable, the $\chi^2$ procedure
gives an unambiguous procedure to determine the best fit assuming that
the data points are statistically independent and have a normally
distributed uncertainty in the dependent variable. When the data
points are presented in a two-dimensional plane with comparable
uncertainties in both axes, there is no unique ``correct'' fitting
procedure (this conclusion holds in both the frequentist and Bayesian
paradigms). In the frequentist picture, the lack of a unique fitting
procedure has led to the use of several methods including orthogonal
least squares, orthogonal regression, and reduced major axis
regression~\cite{Isobe90}. Several of these procedures are often
referred to by different names. Reduced major axis regression is also
referred to as geometric mean regression in Ref.~\cite{Draper97} and a
linear version obtaining the so-called ``impartial line'' was first
used in Ref.~\cite{Stromberg40}. This ambiguity is part of the reason
why more quantitative fits to neutron star cooling data have not
yet been performed before this work. 

We choose to proceed using Bayesian inference, with
\begin{equation}
  P(M|D) \propto P(D|M)P(M)
\end{equation}
where $P(M)$ is the prior distribution for the model $M$ (which in our
case has six parameters for superfluidity or superconductivity and one
parameter for the envelope composition of each star) and $P(D|M)$ is
the likelihood function obtained from the neutron star cooling data.
The quantity $P(M|D)$ is the probability distribution that we want to
obtain, the probability of the theoretical model given the data. In
the Bayesian picture, the non-uniqueness in the fitting procedure
described above is manifest in the undetermined prior distribution
which one must choose to proceed.

Because the uncertainties in the neutron star cooling data are often
presented in terms of the logarithms of temperature and time, we
choose to write the likelihood in terms of new variables $\hat{t}$
and $\hat{T}$,
\begin{eqnarray}
  \widehat{t}&\equiv&\frac{1}{5}\log_{10}
  \left(\frac{t}{10^{2}~\mathrm{yr}}\right)
  \quad \mathrm{and}\nonumber \\
  \widehat{T}&\equiv&\frac{1}{2}\log_{10}
  \left(\frac{T}{10^{5}~\mathrm{K}}\right)
\end{eqnarray}
which are defined so that typical values are between 0 and 1.

We assume that our data set is Gaussian in both variables
$\widehat{t}$ and $\widehat{T}$, and can thus be specified as
$\widehat{t}_j$, $\widehat{T}_j$, $\delta \widehat{t}_j$, and $\delta
\widehat{T}_j$. (Our uncertainties are sufficiently small that the
distinction between normal and log-normal distributions will not
strongly impact our qualitative results.) The
composition of the envelope is parameterized by a quantity
$\eta$ which takes values from 0 to $10^{-7}$, larger values
representing a larger contribution from light elements~\cite{Potekhin97}.
The cooling code computes
three different cooling curves, for $\eta=0$, $\eta=10^{-12}$, and
$\eta=10^{-7}$ and results for other values of $\eta$ are obtained
through linear interpolation. 
The likelihood function is
\begin{eqnarray}
  \cal{L}_{\mathrm{H}} &\propto&
  \prod_{j}{\int d\widehat{t} {\sqrt{\left\{\left[
          \frac{d\widehat{T}(\eta_j,\widehat{t})}
               {d\widehat{t}}\right]^{2}+1\right\}}}}
  \exp\left\{\frac{-\left[\widehat{t}-\widehat{t}_{j}\right]^{2}}
                  {2\left(\delta
                    \widehat{t}_j\right)^{2}}\right\}
                  \nonumber \\
                  && \times\exp\left\{\frac{-\left[\widehat{T}
                      \left(\eta_j,\widehat{t}
                      \right)-\widehat{T}_{j}\right]^{2}}
                            {2\left(\delta
                              \widehat{T}_j\right)^{2}}\right\} \, .
                            \label{eq:like}
\end{eqnarray}
where the product runs over all of the neutron stars in the data set.
The overall normalization is unspecified and is not necessary for our
results. For older neutron stars with spectra well fit by a blackbody
spectrum, $\eta=0$, corresponding to the assumption that a
heavy-element atmosphere implies no light elements in the envelope.
Note that this likelihood function reduces exactly to the likelihood
function for the traditional $\chi^2$ procedure in the limiting cases
that one of the two variables has a small uncertainty. The square root
operates as a line element, specifying how one defines a distance when
integrating the cooling curve along the data. The ambiguity in
defining this distance is the exact same as the choice in using
different frequentist regression techniques. Our approach makes this
ambiguity explicit.

This technique is very similar to the recent determination of the
mass-radius curve given neutron star mass and radius observations (the
formalism was first developed in Ref.~\cite{Steiner10te} and most
recently updated in Ref.~\cite{Steiner16ns}). There are two
significant differences. First, the term under the square root was
ignored, appropriate because the radius depends only very weakly on
the neutron star mass. Second, the data in that case is not Gaussian
in either mass or radius so a more complicated probability
distribution was used rather than the product of two Gaussians
employed in this work.

In practice, the cooling curves are specified as arrays,
$\widehat{T}(\eta,\widehat{t}) \rightarrow
[\widehat{T}_{k}(\eta),\widehat{t}_k]$ and finite differencing
gives the derivative $[d\widehat{T}(\eta)/d\widehat{t}]_k$. To a good
approximation we can replace the integral by a sum
\begin{eqnarray}
  \cal{L} &\propto&
  \prod_{j}{\sum_{k}{\sqrt{\left\{\left[\frac{d\widehat{T}
            (\eta_j)}
          {d\widehat{t}}\right]_k^{2}+1\right\}}}}
  \exp\left\{\frac{-\left[\widehat{t}_k-\widehat{t}_{j}\right]^{2}}
    {2(\delta \widehat{t}_j)^{2}}\right\}
  \nonumber \\
  && \times\exp\left\{\frac{-\left[\widehat{T}_k(\eta_j)-
      \widehat{T}_{j}\right]^{2}}
            {2(\delta \widehat{T}_j)^{2}}\right\} \, .
\end{eqnarray}
over a uniform grid in $\hat{t} \in [0,1]$. In this context, one can
see the purpose for the term under the square root sign: in regions
where the cooling curve is nearly vertical, the data covers fewer grid
points than in regions where the curve is nearly horizontal. The term
under the square root compensates for this, ensuring portions of the
cooling curve which are nearly vertical get extra weight. This
reweighting is relatively weak in comparison to the data, which
exponentially affects the likelihood. We choose a grid of size
100 but increasing the number of grid points will not affect
our basic conclusions. When we fit luminosities rather than
temperatures, we can just replace $\widehat{T}$ with $\widehat{L}
\equiv (1/4) \mathrm{log}_{10} [L/(10^{30}~\mathrm{erg}/\mathrm{s})]$.

The Markov chain Monte Carlo begins with an initial guess for the six
superfluid parameters ($T_{c,\mathrm{peak},n}$,
$k_{F,\mathrm{peak},n}$, $\Delta k_{F,n}$, $T_{c,\mathrm{peak},p}$,
$k_{F,\mathrm{peak},p}$, $\Delta k_{F,p}$) and the envelope
composition parameters. A new set of gaps and envelope compositions is
randomly selected and the new likelihood is computed. The step is
rejected or accepted according to the Metropolis algorithm. The
autocorrelation length of all of the parameters is computed and the
data is block averaged to ensure the uncertainties in the
parameters are properly estimated.

\begin{table*}  
  \begin{tabular}{lccccc}
    Quantity & \multicolumn{3}{c}{Value and 1$-\sigma$ uncertainty} \\
    &  w/o Vela or carbon  &  w/o carbon & All \\
    \hline
   $ \log_{10} T_{c,\mathrm{peak},n}$ &
   $ 8.48 \pm 0.57 $ &
   $ 9.11 \pm 0.19 $ &
   $ 9.61 \pm 0.22 $ \\
   $ k_{F,\mathrm{peak},n}~(\mathrm{fm}^{-1}) $ &
   $ 1.70 \pm 0.12 $ &
   $ 1.924 \pm 0.089 $ &
   $ 1.750 \pm 0.082 $ \\
   $ \Delta k_{F,n}~(\mathrm{fm}^{-1}) $ &
   $ 0.21 \pm 0.09 $ &
   $ 1.980 \pm 0.076 $ &
   $ 1.80  \pm  0.10 $ \\
   $ \log_{10} T_{c,\mathrm{peak},p}$ &
   $ 8.57 \pm 0.60 $ &
   $ 8.81  \pm 0.81 $ &
   $ 8.72 \pm 0.45 $ \\
   $ k_{F,\mathrm{peak},p}~(\mathrm{fm}^{-1}) $ &
   $ 0.67 \pm 0.10 $ &
   $ 0.937  \pm 0.086 $&
   $ 1.02 \pm 0.10 $ \\
   $ \Delta k_{F,p}~(\mathrm{fm}^{-1}) $ &
   $ 0.33  \pm 0.12 $ &
   $ 0.145 \pm 0.061 $ &
   $ 0.219 \pm 0.067 $ \\
   $ \log_{10} \eta_{\mathrm{Cas\: A }}$ &
    &
    &
   $ -9.52  \pm 0.87 $ \\
   $ \log_{10} \eta_{\mathrm{XMMU \: J1732}}$ &
    &
    &
   $ -9.14 \pm 0.70 $ \\
   $ \log_{10} \eta_{\mathrm{PSR \: J1119}}$ &
   $ -14.3 \pm 1.4 $ &
   $ -16.00 \pm 0.77 $&
   $ 16.56 \pm  0.63 $ \\
   $ \log_{10} \eta_{\mathrm{RXJ \: 0822 }}$ & 
   $ -8.80 \pm 0.68 $ & 
   $ -9.80 \pm 0.64 $&
   $ -9.97 \pm 0.84 $ \\
   $ \log_{10} \eta_{\mathrm{ 1E\: 1207 }}$ &
   $ -9.57 \pm 0.87 $ &   
   $ -10.8 \pm 1.0 $&
   $ -10.22 \pm 0.71 \ $ \\
   $ \log_{10} \eta_{\mathrm{ PSR\: J1357 }}$ &
   $ -9.30 \pm 0.81 $ &    
   $ -9.18 \pm 0.50 $&
   $ -9.58 \pm 0.88 $ \\
   $ \log_{10} \eta_{\mathrm{ RX\: J0002 }}$ &
   $ -16.36 \pm 0.68 $ &   
   $ -16.90 \pm  0.39 $&
   $ -16.40 \pm  0.61 $ \\
   $ \log_{10} \eta_{\mathrm{ PSR \: B0833 }}$ &   
   $ $&  
   $ -8.29 \pm 0.49 $&
   $ -8.65 \pm 0.67 $ \\
   $ \log_{10} \eta_{\mathrm{ PSR\: B1706 }}$ &   
   $ -9.35 \pm 0.79 $ & 
   $ -8.9 \pm 1.1 $&
   $ -8.30 \pm 0.51 $ \\
   $ \log_{10} \eta_{\mathrm{ PSR\: J0538 }}$ & 
   $ -16.18 \pm 0.66 $ &   
   $ -16.57 \pm 0.25 $&
   $ -16.88 \pm 0.40 \hphantom{-1} $ \\
   $ \log_{10} \eta_{\mathrm{ PSR \: B2334 }}$ &   
   $ -8.61 \pm 0.79 $ & 
   $ -8.14 \pm 0.32 $ &
   $ -8.12 \pm 0.47 $ \\    
  \end{tabular}
  \caption{Posterior parameter values for three fits of the minimal
    cooling model to data. The first column labels the parameter, the
    second column gives results obtained without including Vela
    (B0833$-$45) or the carbon atmosphere stars, the third column includes
    Vela, and the fourth column includes all of the stars in the data
    set. The gap parameters depend most strongly on
    whether or not Vela is included in the fit. The envelope
    compositions are relatively insensitive to the data selection,
    but vary strongly between individual neutron stars.}
  \label{tab:parmpost}
\end{table*}

\begin{figure*}[htb]
\parbox{0.5\hsize}{
\includegraphics[width=\hsize]{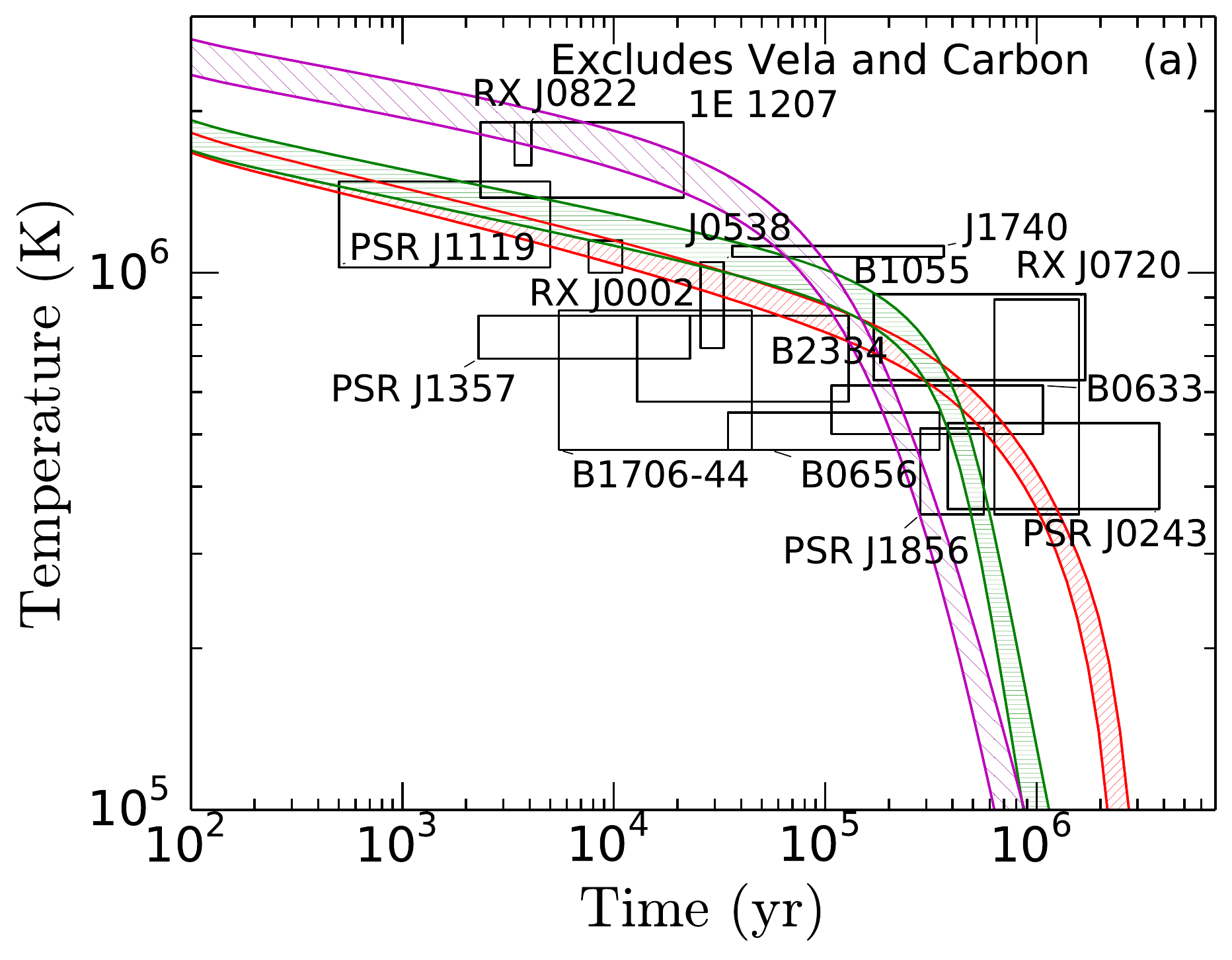}\\[-2ex]
}\parbox{0.5\hsize}{
\includegraphics[width=\hsize]{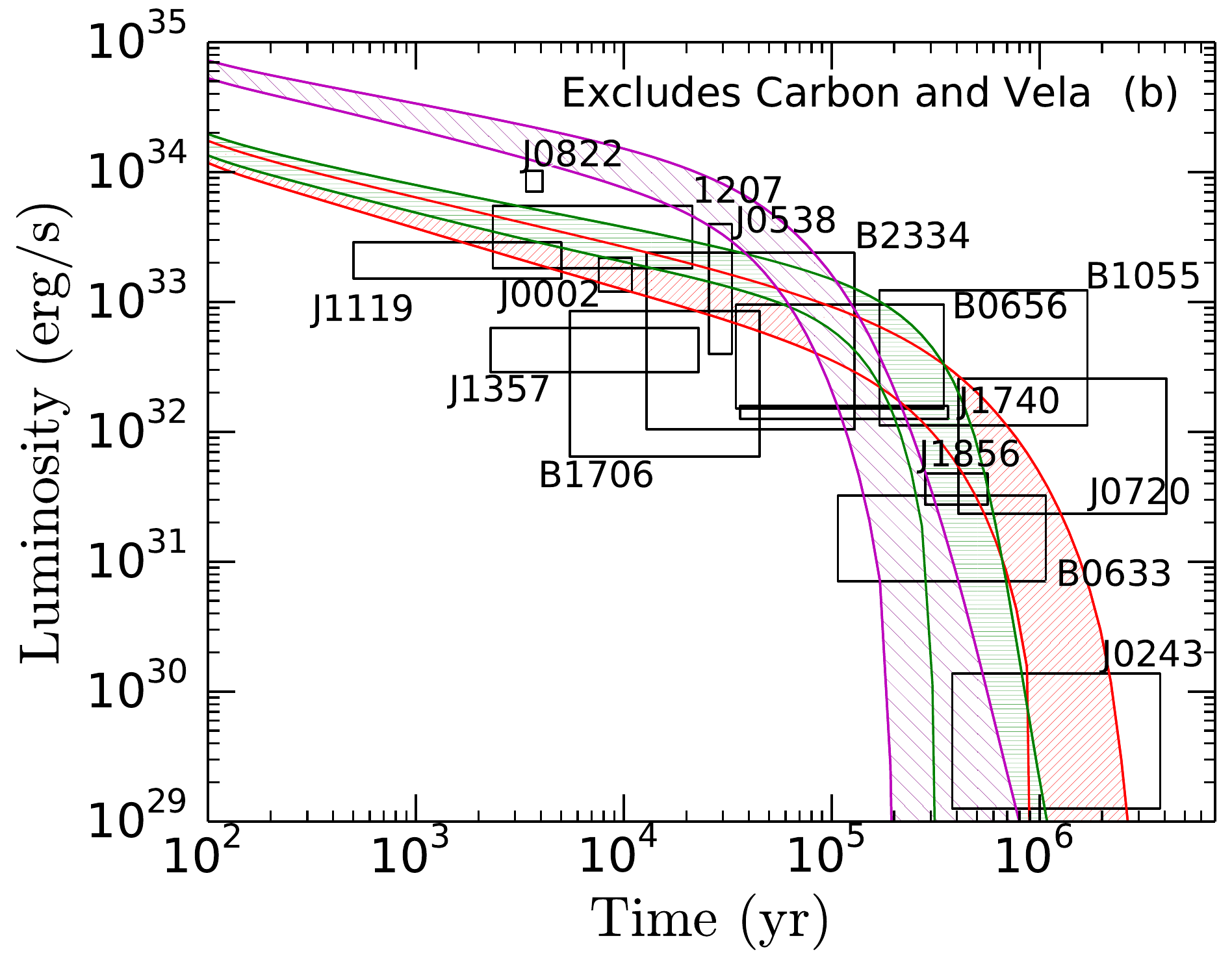}\\[-2ex]
}\\[2ex]
\parbox{0.5\hsize}{
\includegraphics[width=\hsize]{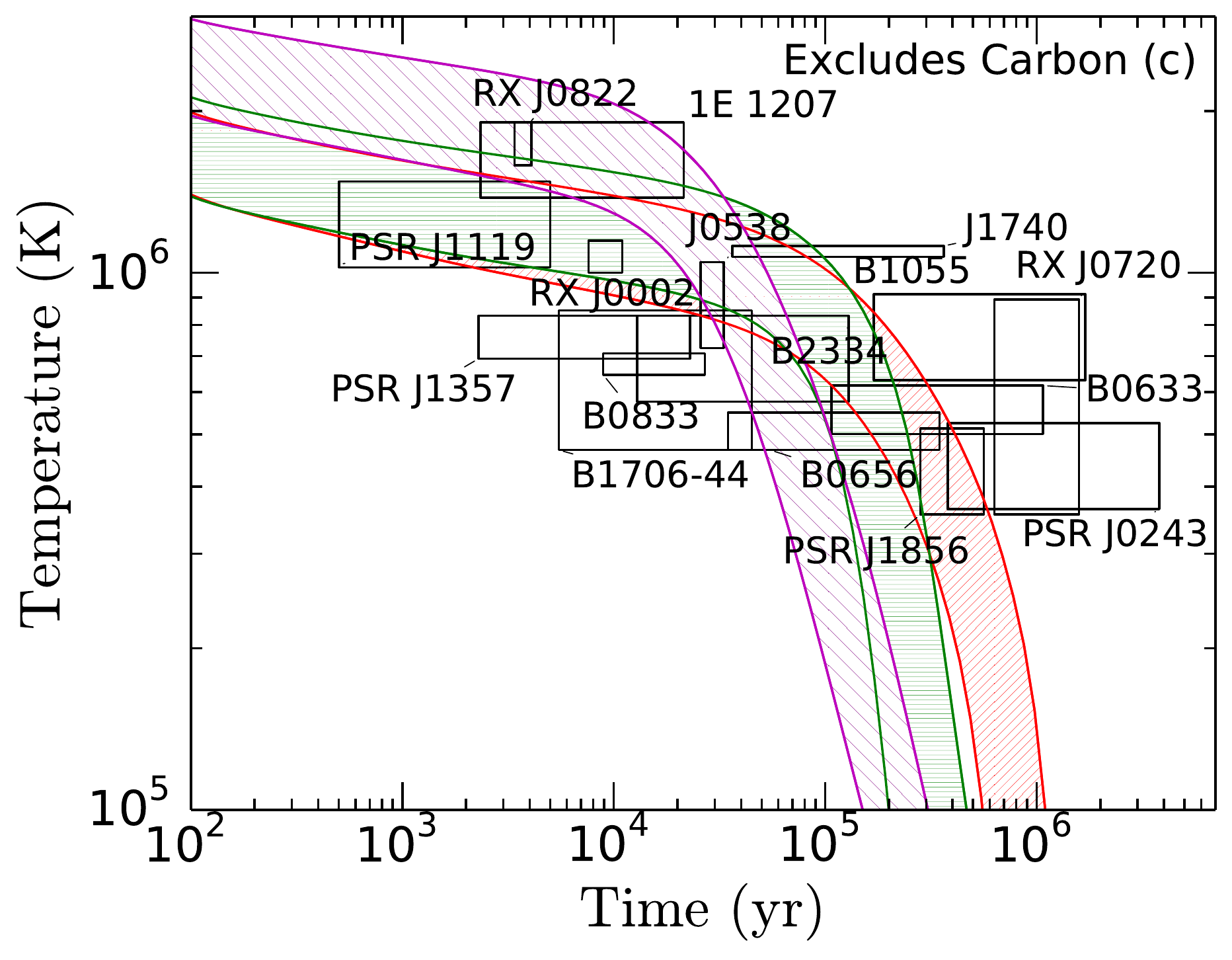}\\[-2ex]
}\parbox{0.5\hsize}{
\includegraphics[width=\hsize]{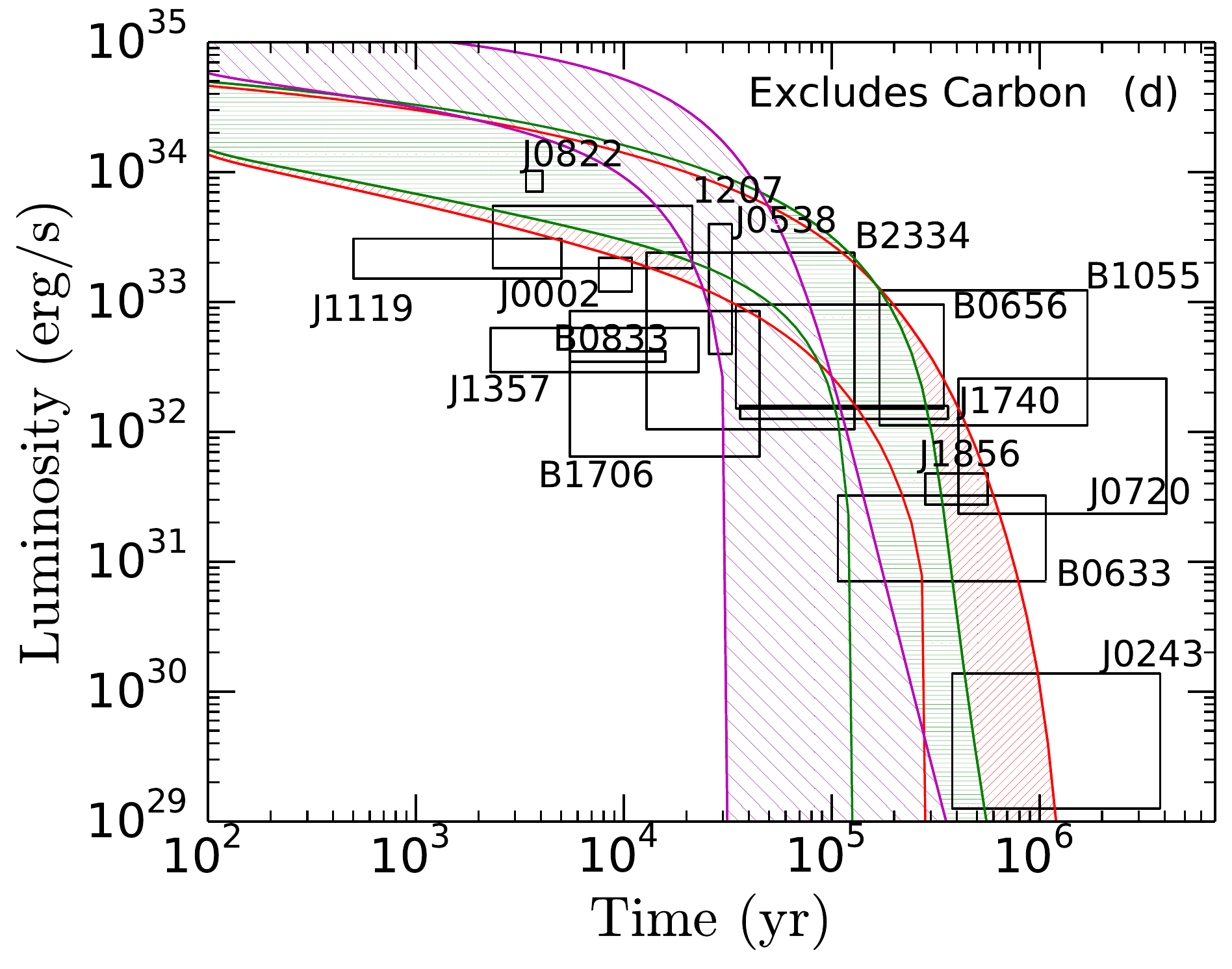}\\[-2ex]
}\\[2ex]
\parbox{0.5\hsize}{
\includegraphics[width=\hsize]{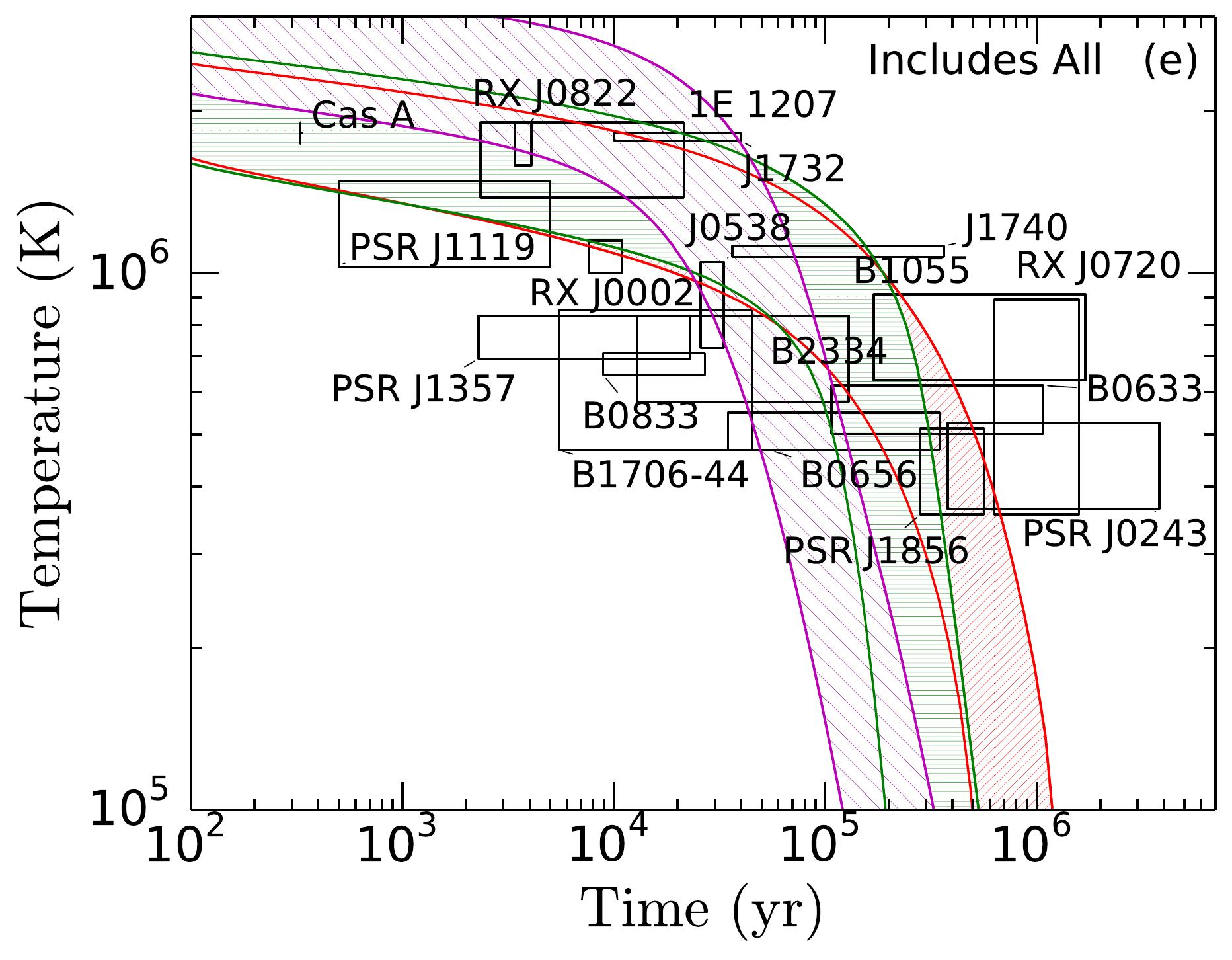}\\[-2ex]
}\parbox{0.5\hsize}{
\includegraphics[width=\hsize]{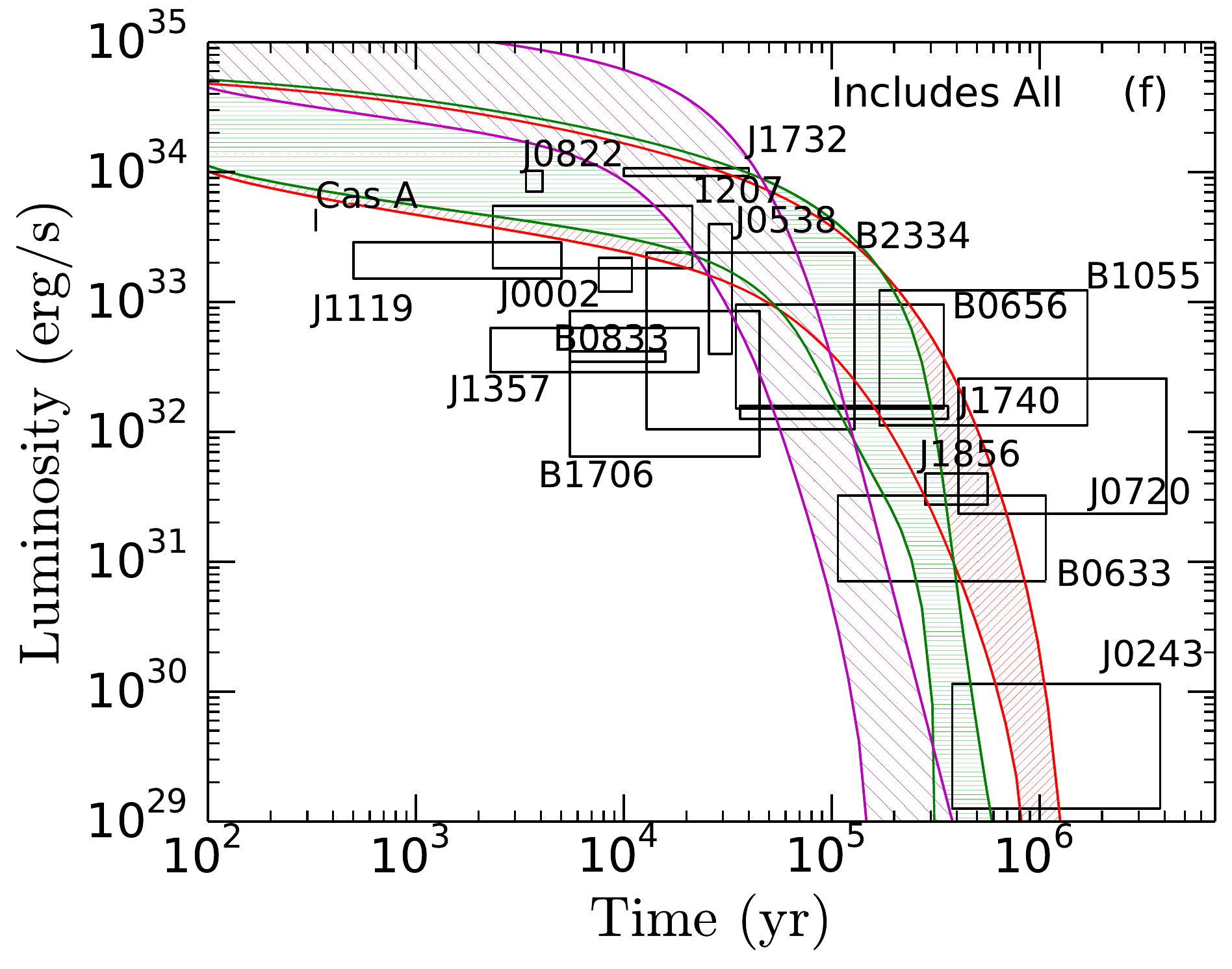}\\[-2ex]
}
\caption{(Color online) Theoretical cooling curves illustrating how
  temperature and luminosity decrease over time. The black boxes
  represent neutron star cooling data and three colored bands show the
  $\pm 1 \sigma$ uncertainties on the cooling curves (from
  superfluidity and superconductivity). The three bands represent
  three different values of $\eta$, $10^{-7}$ (purple, \textbackslash\textbackslash \:  hatching), $10^{-12}$ (green, horizontal hatching)
  and $10^{-17}$ (red, // hatching). The temperature results [labeled (a), (c), and (e)] correspond to the parameter
  limits in Table~\ref{tab:parmpost}, while the luminosity results [labeled (b), (d), and (f)] correspond to the parameter limits in Table~\ref{tab:parmpost2}.}
\label{fig:cool}
\end{figure*}

\begin{figure}
  \includegraphics[width=3.4in]{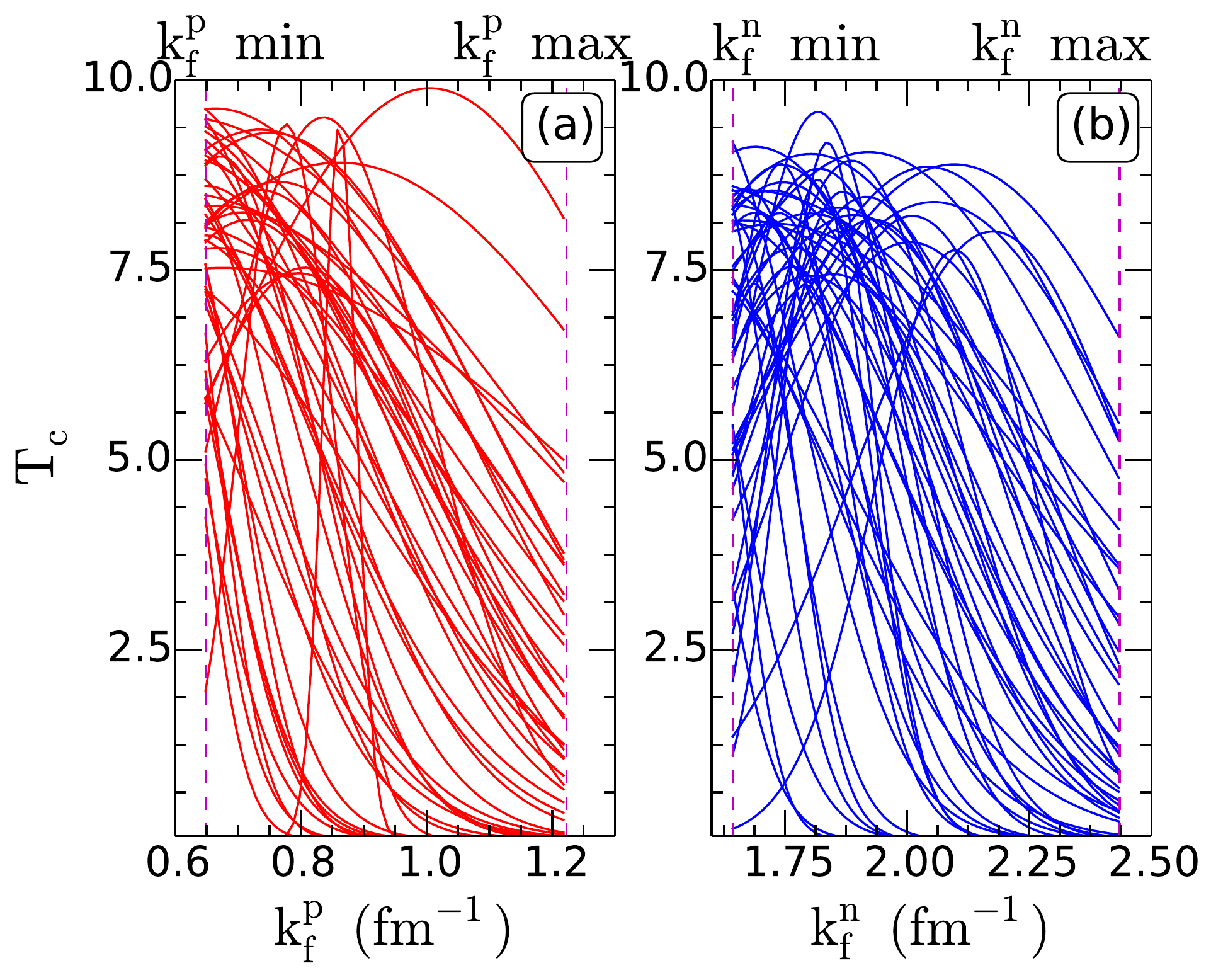}
  \caption{Uncorrelated samples from the critical temperatures
    from the fit to the luminosities
    as a
    function of the Fermi momenta for protons [left panel, (a)] and
    neutrons [right panel, (b)] without Vela or the carbon stars. The left
    boundary in both panels represents the Fermi momentum at the
    crust-core transition (denoted ``$k^{\rm{p}}_{\rm{f}}\, \rm{min}$"). The right
    boundary represents the Fermi momentum in the center (denoted
    ``$k^{\rm{n}}_{\rm{f}}\,\rm{max}$") of a $1.4~\mathrm{M}_{\odot}$ neutron star. The
    uncertainty in the critical temperatures is large and there is a
    preference for proton superconductivity to peak at lower Fermi
    momenta.}
  \label{fig:tc_000}
\end{figure}

\begin{figure}
  \includegraphics[width=3.4in]{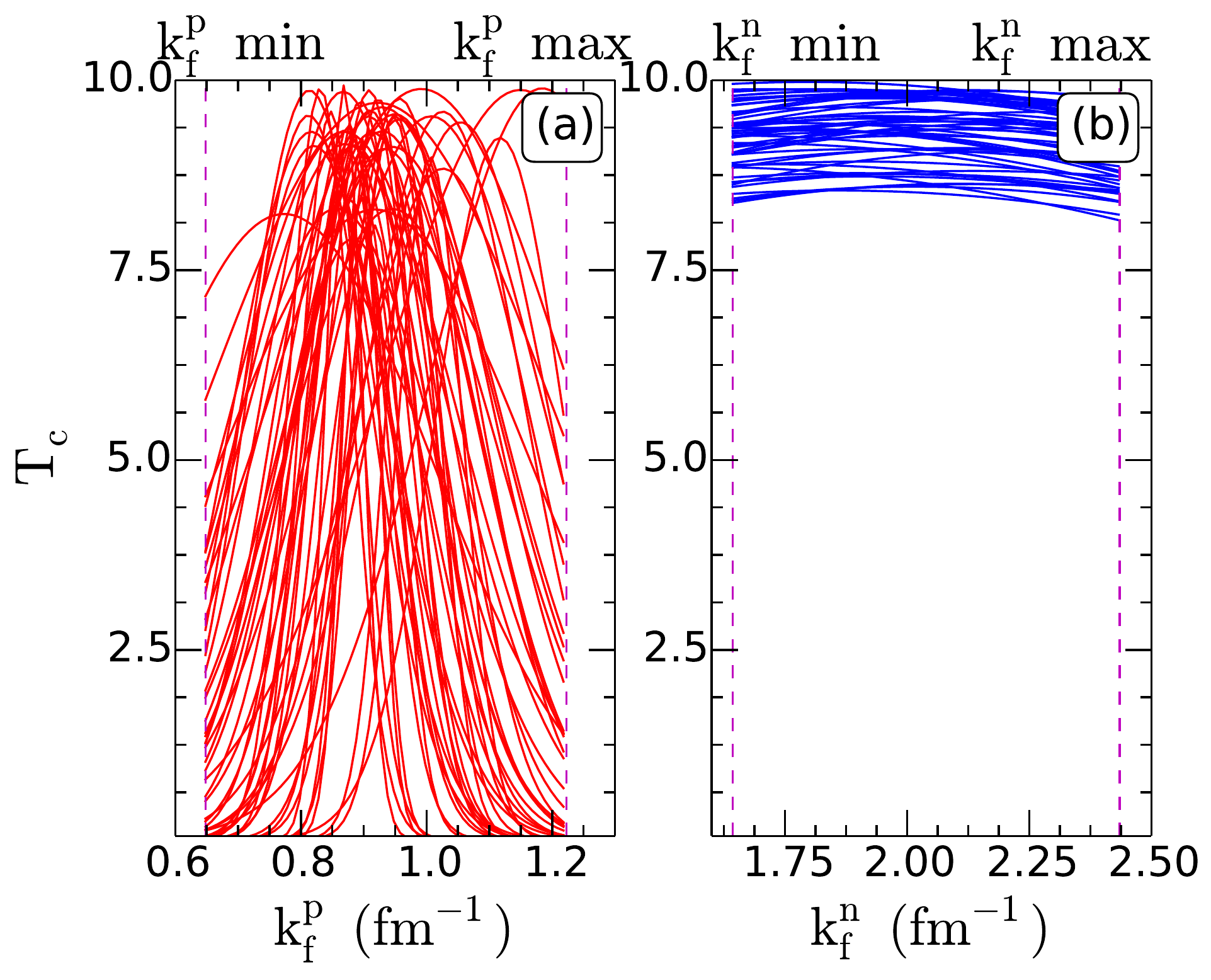}
  \caption{Uncorrelated samples from the critical temperatures from
    the fit to the luminosities as in Fig.~\ref{fig:tc_000} but now
    with Vela. Similar to previous works, we find strong proton
    superconductivity (a) and slightly weaker neutron triplet
    superfluidity  (b). The proton superconductivity moves to higher
    densities and the density dependence of the neutron superfluid gap
    broadens to maximize the cooling to match the low
    luminosity of Vela.}
  \label{fig:tc_001}
\end{figure}

\begin{table*}  
  \begin{tabular}{lccccc}
    Quantity & \multicolumn{3}{c}{Value and 1$-\sigma$ uncertainty} \\
    &  w/o Vela or carbon  &  w/o carbon & all \\
    \hline
    $ \log_{10} T_{c,\mathrm{peak},n}$ &
    $ 8.32 \pm 0.49 $ &
    $ 9.34 \pm 0.38 $ &
    $ 9.55 \pm 0.29 $ \\
    $ k_{F,\mathrm{peak},n}~(\mathrm{fm}^{-1}) $ &
    $ 1.78 \pm 0.16 $ &
    $ 1.95 \pm 0.12  $ &
    $ 1.714 \pm 0.087 $ \\
    $ \Delta k_{F,n}~(\mathrm{fm}^{-1}) $ &
    $ 0.28 \pm 0.12 $ &
    $ 2.05 \pm 0.11  $ &
    $ 1.744 \pm 0.089 $ \\
    $ \log_{10} T_{c,\mathrm{peak},p}$ &
    $ 8.88 \pm 0.63  $ &
    $ 9.26 \pm 0.46 $ &
    $ 8.48 \pm 0.42 $ \\
    $ k_{F,\mathrm{peak},p}~(\mathrm{fm}^{-1}) $ &
    $ 0.641 \pm 0.11 $ &
    $ 0.928 \pm 0.07  $&
    $ 0.99 \pm 0.09 $ \\
    $ \Delta k_{F,p}~(\mathrm{fm}^{-1}) $ &
    $ 0.276 \pm 0.12 $ &
    $ 0.115 \pm 0.047 $ &
    $ 0.21 \pm 0.085 $ \\
    $ \log_{10} \eta_{\mathrm{Cas\: A }}$ &
    &
    &
    $ -9.51  \pm 0.70  $ \\
    $ \log_{10} \eta_{\mathrm{XMMU \: J1732}}$ &
    &
    &
    $ -9.03 \pm 0.81 $ \\
    $ \log_{10} \eta_{\mathrm{PSR \: J1119}}$ &
    $  -14.39 \pm 0.98 $ &
    $ -16.41 \pm 0.40  $&
    $ 16.51 \pm  0.61  $ \\
    $ \log_{10} \eta_{\mathrm{RXJ \: 0822 }}$ & 
    $ -9.3 \pm 1.1 $ & 
    $ -9.88 \pm 0.52 $&
    $ -9.99 \pm 0.64 $ \\
    $ \log_{10} \eta_{\mathrm{ 1E\: 1207 }}$ &
    $ -10.0 \pm 1.1 $ &   
    $ -10.9 \pm 0.52  $&
    $ -10.22 \pm 0.76 \ $ \\
    $ \log_{10} \eta_{\mathrm{ PSR\: J1357 }}$ &
    $ -9.15 \pm 0.94  $ &    
    $ -9.31 \pm 0.68  $&
    $ -9.87 \pm 0.88 $ \\
    $ \log_{10} \eta_{\mathrm{ RX\: J0002 }}$ &
    $ -16.41 \pm 0.80 $ &   
    $ -16.84 \pm  0.39 $&
    $ -16.17 \pm  0.55 $ \\
    $ \log_{10} \eta_{\mathrm{ PSR \: B0833 }}$ &   
    $ $&  
    $ -8.27 \pm 0.38  $&
    $ -8.44 \pm 0.53 $ \\
    $ \log_{10} \eta_{\mathrm{ PSR\: B1706 }}$ &   
    $ -9.46 \pm 1.1  $ & 
    $ -8.33 \pm 0.52  $&
    $ -8.42 \pm 0.50 $ \\
    $ \log_{10} \eta_{\mathrm{ PSR\: J0538 }}$ & 
    $ -16.00 \pm 0.10  $ &   
    $ -16.69 \pm 0.57 $&
    $ -16.80 \pm 0.46 \hphantom{-1} $ \\
    $ \log_{10} \eta_{\mathrm{ PSR \: B2334 }}$ &   
    $ -8.46 \pm 0.64 $ & 
    $ -8.03 \pm 0.35  $ &
    $ -7.88 \pm 0.27 $ \\    
  \end{tabular}
  \caption{Luminosity fits of the same posterior parameter values.}
  \label{tab:parmpost2}
\end{table*}

\section{Results}

We begin by removing Vela (PSR B0833$-$45) and the two carbon
atmosphere stars in Cas A and XMMU J1732. We perform a MCMC simulation
as described above, assuming that the minimal cooling model holds,
i.e., that the direct Urca process does not operate and that no exotic
matter is present. We fit the theoretical effective surface
temperatures (accounting for the redshift factor) to the temperatures
in Table~\ref{tab:data} implied by the x-ray spectra. The resulting
gap parameters, the envelope compositions, and their uncertainties
(which we have assumed symmetric with respect to their central values)
are given in the first column of Table~\ref{tab:parmpost}. The
posterior cooling curves for $\eta=0, 10^{-12}$, and $10^{-7}$ are
plotted in the top left panel of Fig.~\ref{fig:cool}. We find large
uncertainties in the critical temperatures for the singlet proton gap
and the triplet neutron gap, and our numbers are not in disagreement
with previous results from Ref.~\cite{Page09ne}.

The results from fitting the luminosities rather than the temperatures
are presented in the upper right panel of Fig.~\ref{fig:cool} and the
first column of Table~\ref{tab:parmpost2}. The results are relatively
similar to those obtained by fitting the temperature rather than the
luminosity. Representative curves which show the dependence of the
superfluid gaps on Fermi momentum are given in Fig.~\ref{fig:tc_000},
showing that the proton superconducting gap is likely largest just
near the crust core transition and falls off dramatically at the
highest densities in the core. The triplet neutron superfluid critical
temperature, on the other hand, may peak at any density so long as a
large enough portion of the core undergoes the superfluid phase
transition.

The quantitative nature of our fit also allows us to determine the
envelope composition for H atmosphere neutron stars. We find PSR
J1119-6127, RX J0002+6246 and PSR J0538+2817 all most likely have no
light elements in their envelopes, in contrast with a small amount of
light elements in 1E 1207.4-5209 and a significant contribution from
light elements in all of the other H atmosphere stars. Note that stars
which lie to the left and below the cooling curves tend to have a
large amount of light elements, fitting better to the $\eta=10^{-7}$
(purple) curve lying to the right of the data point than
the $\eta=10^{-17}$ (red) curve above the data point (because
the time uncertainty is larger than the temperature uncertainty).

Now we add Vela and redo the temperature fit. The results are
summarized in the second column of Table~\ref{tab:parmpost}, and the
middle left panel of Fig.~\ref{fig:cool}. This one data point,
lying to the left and below the curves, has a strong impact: The
critical temperatures implied by the data are much larger than those
obtained previously. We find neutron superfluid critical temperatures
near $10^{9}$ K are required to explain the data and the width of the
Gaussian increases significantly allowing a large part of the core to
participate in the Cooper pair neutrino emissivity. The proton superconducting
gap also increases slightly and moves to higher densities. The fit to
the luminosities shown in the second column of
Table~\ref{tab:parmpost2} and the middle right panel in
Fig.~\ref{fig:cool} shows the same trend. Representative curves which
show the critical temperature are given in Fig.~\ref{fig:tc_001}. The
increase in gaps leads to a larger uncertainty in the cooling curves,
as a larger part of the star now participates in the pair-breaking
neutrino emissivity and thus the cooling is more sensitive to the
gaps. The dramatic effect of Vela is partially because of the age revision
of Vela down to $(5-16) \times 10^{3}$ years as obtained in
Ref.~\cite{Tsuruta09} and discussed in Ref.~\cite{Page09ne}. The
envelope compositions are unchanged (within errors) and the fit
prefers a significant amount of light elements in Vela's envelope 
to become closer to the $\eta=10^{-7}$ curve lying to the right.

While the absolute normalization of the likelihood function is not
meaningful, relative values are physical. A typical data point
contributes a factor of 0.5 to the likelihood while Vela's
contribution is $10^{-3}$. This is a strong indication that fitting
Vela is difficult in the minimal cooling model. The observation of
Vela, as it currently stands, provides some evidence for the direct
Urca process or the presence of exotic matter in neutron star cores.

\begin{figure}
  \includegraphics[width=3.4in]{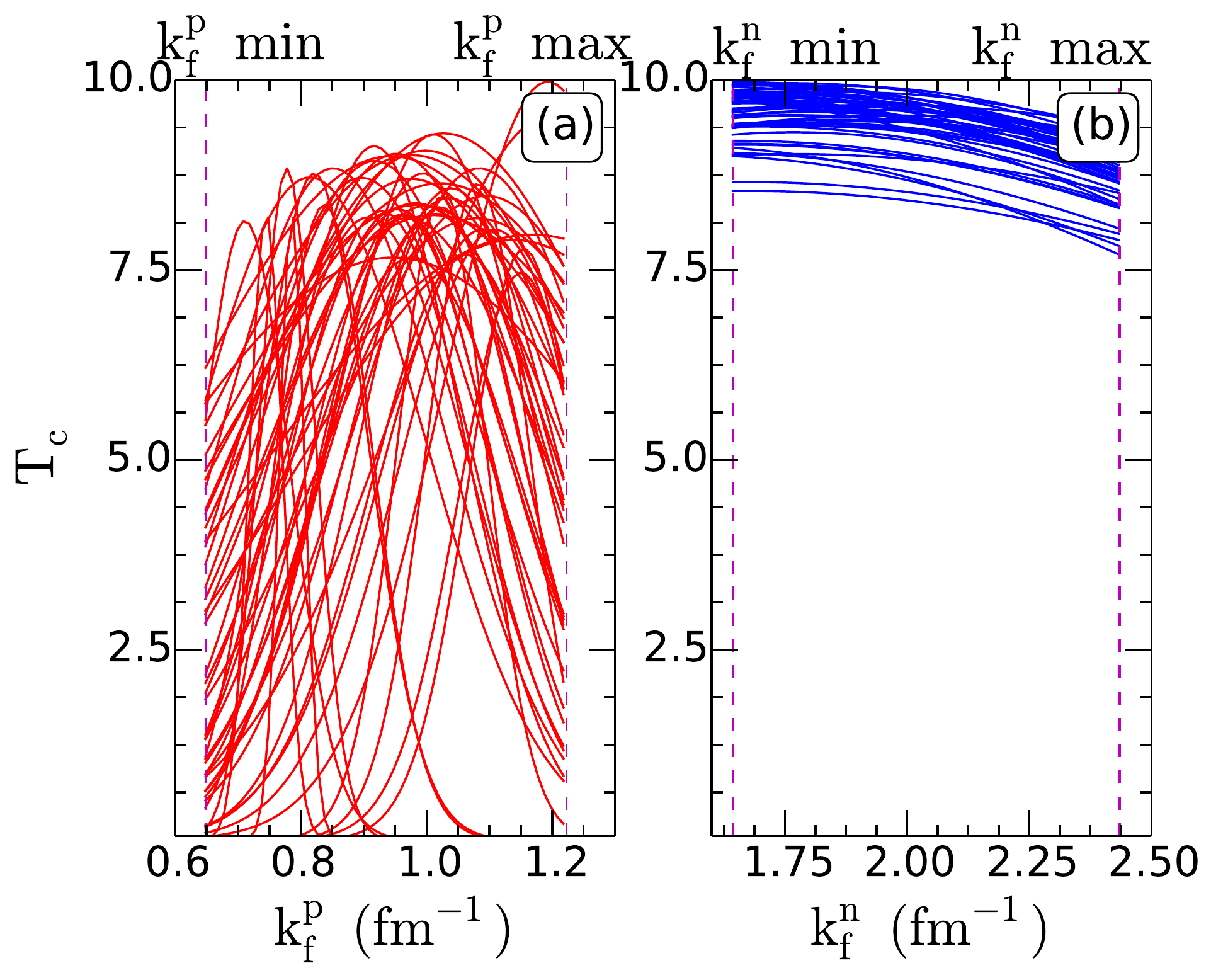}
  \caption{Uncorrelated samples from the critical temperatures
    from the fit to the luminosities as in Figs.~\ref{fig:tc_000} and
    \ref{fig:tc_001} now having added both Vela (B0833$-$45) and the
    carbon-atmosphere stars to the analysis [(a) displays
     proton critical temperatures; (b)
     displays neutron critical temperatures]. Comparing to
    Fig.~\ref{fig:tc_001}, the proton critical temperatures are
    smaller. The smaller proton critical temperature ensures younger
    stars are warm enough to match the relatively large luminosities
    from these two stars.}
  \label{fig:tc_111}
\end{figure}

Previous works~\cite{Page11rc,Shternin11} found very strong
constraints on proton singlet superfluidity and neutron triplet
superfluidity from observations which implied the neutron star in Cas
A had cooled over a 10-year observation period~\cite{Ho09}.
References~\cite{Blaschke12,Blaschke13} present an alternative explanation:
In-medium effects on thermal conductivity as well as the presence of a
particular proton gap explain the cooling. Reference~\cite{Ho15} found
similar constraints on the gaps as found in
Refs.~\cite{Page11rc,Shternin11}, and employed a polynomial
parametrization of the gaps [in contrast to the Gaussian form we use
in Eq.(~\ref{eq:gapparm})]. Recent observations of the neutron star in
Cas A imply that it may not have cooled appreciably in the past 15
years~\cite{Elshamouty13,Posselt13}. For this work, we assume that the
systematics do not enable us to constrain the cooling over a short
time scale.

Employing this assumption, adding the neutron star in Cas A to the
data set does not make a strong modification in our results. Because the
surface temperature of Cas A lies in between the results for envelopes
with and without light elements we simply choose a
moderate amount of light elements $\eta\sim 10^{-10}$ to
explain the data. However, adding the other neutron star thought to
have a carbon atmosphere, XMMU J1732, creates a strong preference for
warmer stars with light element envelopes. The results are summarized in 
the third column of Table~\ref{tab:parmpost} (for the temperature
fit), and the third column of Table~\ref{tab:parmpost2} (for the
luminosity fit), and the bottom panels of Figs.~\ref{fig:cool} and
~\ref{fig:tc_111}. We find strong neutron superfluidity is
required with a weak dependence on the neutron Fermi momenta and
moderate proton superfluidity with a larger uncertainty on the proton
Fermi momentum for which the critical temperature is maximized. These
results are in strong tension with Vela, which has a strong preference
for cooler stars with light element envelopes. This tension results in
very tight constraints on the superfluid properties of dense matter.
In the context of Bayesian inference where the evidence for a
particular model is determined by the integral over the likelihood,
the dramatic decrease in the parameter uncertainties leads to a model
with very small evidence. In other words, if Vela and XMMU J1732 are
confirmed to have ages and temperatures near the central values
reported in Table~\ref{tab:data}, then it is likely that a model with
some additional parameter which enables faster cooling in Vela will
provide a much better fit.

\section{Discussion} 

Most importantly, our work {\em quantifies} the extent to which
superfluid properties can be constrained from currently available data
on the cooling of isolated neutron stars. Most of the previous works
on this topic give more qualitative results: They do not employ any
particular likelihood function and thus cannot give full posteriors
for their parameter values. The extent to which our quantitative
approach will be possible without making the assumptions of the
minimal cooling model will be explored in future work.

Our analysis has either 14, 15, or 17 parameters corresponding to 15,
16, or 18 data points, respectively. One of the advantages of our
Bayesian approach is that our formalism does not require the fitting
problem to be strongly over-constrained. Had we not employed the
minimal cooling model, we would have required at least four new
parameters to describe the EOS and an additional mass parameter for
each neutron star (bringing us to a total of 39 parameters for 18 data
points). An accurate mass measurement for even a few of the neutron
stars in this data set would improve the fitting problem
substantially.

One possible extension would be to attempt to explain the surface
temperatures of accreting neutron stars as well, as done in
Refs.~\cite{Beznogov15a} and \cite{Beznogov15b}. It is well known that
some of those objects, in particular SAX J1808.4$-$3658, are too cold
to be explained within the minimal cooling model~\cite{Heinke07}, and
thus the direct Urca process is invoked. The approach taken in
Refs.~\cite{Beznogov15a} and \cite{Beznogov15b} is similar in that
they employ a systematic exploration of their parameter space; it is
different in that they do not explicitly compute the likelihood of
their models as we have done in Eq.~(\ref{eq:like}). Extending our
method to include the direct Urca process would necessitate also
considering the variation in the EOS as well.

Our theoretical model presumes that the surface temperature of the
neutron star does not vary across the surface. Hot spots on the
neutron star surface may not create pulsations in the emission if they
lie near the axis of rotation. It was argued that fits to the
luminosity rather than the effective temperature partially ameliorate
this difficulty because uneven temperature distributions impact the
shape of the spectrum more strongly than the
luminosity~\cite{Potekhin:2015qsa}. Our results demonstrate that the
luminosity and temperature fits obtain qualitatively similar
constraints on the superfluid gaps with some quantitative differences
(for example, the luminosity fit implies different critical
temperatures for proton superconductivity, especially when Vela is
included). Nevertheless, fitting to luminosities rather than
temperatures may be insufficient to fully explain the data if the
temperature variation across the surface is dramatic.

Our model computes an effective surface temperature based on an
atmosphere model and the amount of light elements in the envelope (see
Ref.~\cite{Potekhin:2014hja} for a recent review). The observed x-ray
data is analyzed presuming a H atmosphere (sometimes including
an estimate of the magnetic field), a carbon atmosphere, or a black body
spectrum. Our results are thus limited by these two ingredients
insofar as they allow us to correctly determine the temperature at the
base of the envelope. 

Several authors have examined the cooling of isolated neutron stars
outside the minimal model. Reference~\cite{Tsuruta09} examined cooling with
hyperons, and finds that superfluidity is required to ensure that the
direct Urca process does not make neutron stars too cold. By allowing
the direct Urca process,
Refs.~\cite{deCarvalho15,Grigorian15,Grigorian16,Sedrakian16} obtain a
strong EOS dependence in their results. These works, along with
Refs.~\cite{Beznogov15a,Beznogov15b}, find that the data can be
explained without exotic matter so long as the direct Urca process
operates in some stars. We find (as first found in
Ref.~\cite{Page04}), that the isolated neutron stars (with the
exception of the Vela pulsar) can be easily explained without having
to invoke the direct Urca process, so long as one allows for
variations in the envelope composition at early times.
Reference~\cite{Sedrakian16b} has invoked axions in a model which does not
include the direct Urca process. While we are performing our work in a
model which contains more restrictive assumptions about the nature of
dense matter, our statistical analysis allows us to be more
quantitative in our conclusions. Extensions of this work beyond the
minimal cooling model are in progress.

For the neutron stars with a carbon atmosphere, Ref.~\cite{Ho15}
performs a $\chi^2$ fit to the data for the neutron star in Cas A,
under the alternative assumption that this neutron star is indeed
cooling quickly as found in Ref.~\cite{Ho09}. A $\chi^2$ fit is
possible here because there is no uncertainty in the $x$ axis, and thus
the likelihood function in Eq.~(\ref{eq:like}) gives the same result. We
include a larger data set and perform our Monte Carlo over a much
larger set of cooling models. Reference~\cite{Noda13} also assumes that Cas
A is cooling quickly, and explains the data using a neutrino
emissivity from superconducting quarks. The cooling of the carbon
atmosphere star XMMU J1732 was addressed in
Ref.~\cite{Ofengeim15}, who also found a large heat blanketing envelope
was required to reproduce the data. Reference~\cite{Ofengeim15}
also obtained a constraint on the
mass and radius of this neutron star because, in their model, the
proton superfluid gap is correlated with the mass and radius. In
contrast, we treat the EOS and superfluid properties of matter as
independent. Ref.~\cite{Suleimanov:2017brq} has argued that the x-ray
spectra of Cas A and XMMU J1732 can also be modeled as H
atmospheres with hot spots as opposed to uniformly emitting carbon
modeled surfaces. This possibility will be considered in future work.

We have presented results with and without Vela, the neutron star in
Cas A, and XMMU J1732, but we cannot yet definitively determine
whether or not those objects should be included or left out. The
decrease in the fit quality may support going beyond the minimal model
to explain Vela and an alternative interpretation for XMMU J1732 (such
as that in Ref.~\cite{Suleimanov:2017brq}), but the final answer
on this question requires more data or smaller uncertainties.

\section{Acknowledgements} 

The authors would like to thank Jim Lattimer and Madappa Prakash for
useful discussions. S.B., S.H., and A.W.S. were supported by Grant No. NSF
PHY 1554876. This work was supported by U.S. DOE Office of Nuclear
Physics. This project used computational resources from the University
of Tennessee and Oak Ridge National Laboratory's Joint Institute for
Computational Sciences.

\bibliographystyle{apsrev}
\bibliography{paper}

\begin{thebibliography}{82}
\expandafter\ifx\csname natexlab\endcsname\relax\def\natexlab#1{#1}\fi
\expandafter\ifx\csname bibnamefont\endcsname\relax
  \def\bibnamefont#1{#1}\fi
\expandafter\ifx\csname bibfnamefont\endcsname\relax
  \def\bibfnamefont#1{#1}\fi
\expandafter\ifx\csname citenamefont\endcsname\relax
  \def\citenamefont#1{#1}\fi
\expandafter\ifx\csname url\endcsname\relax
  \def\url#1{\texttt{#1}}\fi
\expandafter\ifx\csname urlprefix\endcsname\relax\def\urlprefix{URL }\fi
\providecommand{\bibinfo}[2]{#2}
\providecommand{\eprint}[2][]{\url{#2}}

\bibitem[{\citenamefont{Lattimer and Prakash}(2001)}]{Lattimer01}
\bibinfo{author}{\bibfnamefont{J.~M.} \bibnamefont{Lattimer}} \bibnamefont{and}
  \bibinfo{author}{\bibfnamefont{M.}~\bibnamefont{Prakash}},
  \bibinfo{journal}{Astrophys. J.} \textbf{\bibinfo{volume}{550}},
  \bibinfo{pages}{426} (\bibinfo{year}{2001}),
  \urlprefix\url{http://dx.doi.org/10.1086/319702}.

\bibitem[{\citenamefont{Steiner et~al.}(2015)\citenamefont{Steiner, Gandolfi,
  Fattoyev, and Newton}}]{Steiner15un}
\bibinfo{author}{\bibfnamefont{A.~W.} \bibnamefont{Steiner}},
  \bibinfo{author}{\bibfnamefont{S.}~\bibnamefont{Gandolfi}},
  \bibinfo{author}{\bibfnamefont{F.~J.} \bibnamefont{Fattoyev}},
  \bibnamefont{and} \bibinfo{author}{\bibfnamefont{W.~G.}
  \bibnamefont{Newton}}, \bibinfo{journal}{Phys. Rev. C}
  \textbf{\bibinfo{volume}{91}}, \bibinfo{pages}{015804}
  (\bibinfo{year}{2015}),
  \urlprefix\url{http://dx.doi.org/10.1103/PhysRevC.91.015804}.

\bibitem[{\citenamefont{N\"{a}ttil\"{a}
  et~al.}(2016)\citenamefont{N\"{a}ttil\"{a}, Steiner, Kajava, Suleimanov, and
  Poutanen}}]{Nattila16eo}
\bibinfo{author}{\bibfnamefont{J.}~\bibnamefont{N\"{a}ttil\"{a}}},
  \bibinfo{author}{\bibfnamefont{A.~W.} \bibnamefont{Steiner}},
  \bibinfo{author}{\bibfnamefont{J.~J.~E.} \bibnamefont{Kajava}},
  \bibinfo{author}{\bibfnamefont{V.~F.} \bibnamefont{Suleimanov}},
  \bibnamefont{and} \bibinfo{author}{\bibfnamefont{J.}~\bibnamefont{Poutanen}},
  \bibinfo{journal}{Astron. Astrophys.} \textbf{\bibinfo{volume}{591}},
  \bibinfo{pages}{A25} (\bibinfo{year}{2016}),
  \urlprefix\url{http://dx.doi.org/10.1051/0004-6361/201527416}.

\bibitem[{\citenamefont{Ozel and Freire}(2016)}]{Ozel16}
\bibinfo{author}{\bibfnamefont{F.}~\bibnamefont{Ozel}} \bibnamefont{and}
  \bibinfo{author}{\bibfnamefont{P.}~\bibnamefont{Freire}},
  \bibinfo{journal}{Ann. Rev. Astron. Astrophys.}
  \textbf{\bibinfo{volume}{54}}, \bibinfo{pages}{401} (\bibinfo{year}{2016}),
  \urlprefix\url{http://dx.doi.org/10.1146/annurev-astro-081915-023322}.

\bibitem[{\citenamefont{Gandolfi et~al.}(2015)\citenamefont{Gandolfi, Gezerlis,
  and Carlson}}]{Gandolfi15nm}
\bibinfo{author}{\bibfnamefont{S.}~\bibnamefont{Gandolfi}},
  \bibinfo{author}{\bibfnamefont{A.}~\bibnamefont{Gezerlis}}, \bibnamefont{and}
  \bibinfo{author}{\bibfnamefont{J.}~\bibnamefont{Carlson}},
  \bibinfo{journal}{Ann. Rev. Nucl. Part. Sci.} \textbf{\bibinfo{volume}{65}},
  \bibinfo{pages}{303} (\bibinfo{year}{2015}),
  \urlprefix\url{http://dx.doi.org/10.1146/annurev-nucl-102014-021957}.

\bibitem[{\citenamefont{Page et~al.}(2014)\citenamefont{Page, Lattimer,
  Prakash, and Steiner}}]{Page14ss}
\bibinfo{author}{\bibfnamefont{D.}~\bibnamefont{Page}},
  \bibinfo{author}{\bibfnamefont{J.~M.} \bibnamefont{Lattimer}},
  \bibinfo{author}{\bibfnamefont{M.}~\bibnamefont{Prakash}}, \bibnamefont{and}
  \bibinfo{author}{\bibfnamefont{A.~W.} \bibnamefont{Steiner}},
  \emph{\bibinfo{title}{Stellar Superfluids}} (\bibinfo{year}{2014}),
  chap.~\bibinfo{chapter}{21}, ISBN \bibinfo{isbn}{9780198719267},
  \urlprefix\url{http://www.arxiv.org/abs/1302.6626}.

\bibitem[{\citenamefont{Page et~al.}(2004)\citenamefont{Page, Lattimer,
  Prakash, and Steiner}}]{Page04}
\bibinfo{author}{\bibfnamefont{D.}~\bibnamefont{Page}},
  \bibinfo{author}{\bibfnamefont{J.~M.} \bibnamefont{Lattimer}},
  \bibinfo{author}{\bibfnamefont{M.}~\bibnamefont{Prakash}}, \bibnamefont{and}
  \bibinfo{author}{\bibfnamefont{A.~W.} \bibnamefont{Steiner}},
  \bibinfo{journal}{Astrophys. J. Supp.} \textbf{\bibinfo{volume}{155}},
  \bibinfo{pages}{623} (\bibinfo{year}{2004}),
  \urlprefix\url{http://dx.doi.org/10.1086/424844}.

\bibitem[{\citenamefont{Yakovlev and Pethick}(2004)}]{Yakovlev04}
\bibinfo{author}{\bibfnamefont{D.}~\bibnamefont{Yakovlev}} \bibnamefont{and}
  \bibinfo{author}{\bibfnamefont{C.}~\bibnamefont{Pethick}},
  \bibinfo{journal}{Annual Review of Astronomy and Astrophysics}
  \textbf{\bibinfo{volume}{42}}, \bibinfo{pages}{169} (\bibinfo{year}{2004}),
  \urlprefix\url{http://dx.doi.org/10.1146/annurev.astro.42.053102.134013}.

\bibitem[{\citenamefont{Page and Reddy}(2006)}]{Page06}
\bibinfo{author}{\bibfnamefont{D.}~\bibnamefont{Page}} \bibnamefont{and}
  \bibinfo{author}{\bibfnamefont{S.}~\bibnamefont{Reddy}},
  \bibinfo{journal}{Ann. Rev. Nucl. Part. Sci.} \textbf{\bibinfo{volume}{56}},
  \bibinfo{pages}{327} (\bibinfo{year}{2006}),
  \urlprefix\url{http://dx.doi.org/10.1146/annurev.nucl.56.080805.140600}.

\bibitem[{\citenamefont{Steiner et~al.}(2016)\citenamefont{Steiner, Lattimer,
  and Brown}}]{Steiner16ns}
\bibinfo{author}{\bibfnamefont{A.~W.} \bibnamefont{Steiner}},
  \bibinfo{author}{\bibfnamefont{J.~M.} \bibnamefont{Lattimer}},
  \bibnamefont{and} \bibinfo{author}{\bibfnamefont{E.~F.} \bibnamefont{Brown}},
  \bibinfo{journal}{Eur. Phys. J. A} \textbf{\bibinfo{volume}{52}},
  \bibinfo{pages}{18} (\bibinfo{year}{2016}),
  \urlprefix\url{http://dx.doi.org/10.1140/epja/i2016-16018-1}.

\bibitem[{\citenamefont{{Gudmundsson} et~al.}(1983)\citenamefont{{Gudmundsson},
  {Pethick}, and {Epstein}}}]{Gudmundsson83}
\bibinfo{author}{\bibfnamefont{E.~H.} \bibnamefont{{Gudmundsson}}},
  \bibinfo{author}{\bibfnamefont{C.~J.} \bibnamefont{{Pethick}}},
  \bibnamefont{and} \bibinfo{author}{\bibfnamefont{R.~I.}
  \bibnamefont{{Epstein}}}, \bibinfo{journal}{Astrophys. J.}
  \textbf{\bibinfo{volume}{272}}, \bibinfo{pages}{286} (\bibinfo{year}{1983}),
  \urlprefix\url{https://dx.doi.org/10.1086/161292}.

\bibitem[{\citenamefont{Potekhin et~al.}(1997)\citenamefont{Potekhin, Chabrier,
  and Yakovlev}}]{Potekhin97}
\bibinfo{author}{\bibfnamefont{A.~Y.} \bibnamefont{Potekhin}},
  \bibinfo{author}{\bibfnamefont{G.}~\bibnamefont{Chabrier}}, \bibnamefont{and}
  \bibinfo{author}{\bibfnamefont{D.~G.} \bibnamefont{Yakovlev}},
  \bibinfo{journal}{Astron. Astrophys.} \textbf{\bibinfo{volume}{323}},
  \bibinfo{pages}{415} (\bibinfo{year}{1997}),
  \urlprefix\url{http://aa.springer.de/papers/7323002/2300415.pdf}.

\bibitem[{\citenamefont{Potekhin}(2014)}]{Potekhin:2014hja}
\bibinfo{author}{\bibfnamefont{A.~Y.} \bibnamefont{Potekhin}},
  \bibinfo{journal}{Phys. Usp.} \textbf{\bibinfo{volume}{57}},
  \bibinfo{pages}{735} (\bibinfo{year}{2014}),
  \urlprefix\url{https://dx.doi.org/10.3367/UFNe.0184.201408a.0793}.

\bibitem[{\citenamefont{{Page} et~al.}(2009)\citenamefont{{Page}, {Lattimer},
  {Prakash}, and {Steiner}}}]{Page09ne}
\bibinfo{author}{\bibfnamefont{D.}~\bibnamefont{{Page}}},
  \bibinfo{author}{\bibfnamefont{J.~M.} \bibnamefont{{Lattimer}}},
  \bibinfo{author}{\bibfnamefont{M.}~\bibnamefont{{Prakash}}},
  \bibnamefont{and} \bibinfo{author}{\bibfnamefont{A.~W.}
  \bibnamefont{{Steiner}}}, \bibinfo{journal}{Astrophys. J.}
  \textbf{\bibinfo{volume}{707}}, \bibinfo{pages}{1131} (\bibinfo{year}{2009}),
  \urlprefix\url{http://dx.doi.org10.1088/0004-637X/707/2/1131}.

\bibitem[{\citenamefont{Fesen et~al.}(2006)\citenamefont{Fesen, Hammell, Morse,
  Chevalier, Borkowski, Dopita, Gerardy, Lawrence, Raymond, and van~den
  Bergh}}]{Fesen:2006zma}
\bibinfo{author}{\bibfnamefont{R.~A.} \bibnamefont{Fesen}},
  \bibinfo{author}{\bibfnamefont{M.~C.} \bibnamefont{Hammell}},
  \bibinfo{author}{\bibfnamefont{J.}~\bibnamefont{Morse}},
  \bibinfo{author}{\bibfnamefont{R.~A.} \bibnamefont{Chevalier}},
  \bibinfo{author}{\bibfnamefont{K.~J.} \bibnamefont{Borkowski}},
  \bibinfo{author}{\bibfnamefont{M.~A.} \bibnamefont{Dopita}},
  \bibinfo{author}{\bibfnamefont{C.~L.} \bibnamefont{Gerardy}},
  \bibinfo{author}{\bibfnamefont{S.~S.} \bibnamefont{Lawrence}},
  \bibinfo{author}{\bibfnamefont{J.~C.} \bibnamefont{Raymond}},
  \bibnamefont{and} \bibinfo{author}{\bibfnamefont{S.}~\bibnamefont{van~den
  Bergh}}, \bibinfo{journal}{Astrophys. J.} \textbf{\bibinfo{volume}{645}},
  \bibinfo{pages}{283} (\bibinfo{year}{2006}),
  \urlprefix\url{https://dx.doi.org/10.1086/504254}.

\bibitem[{\citenamefont{Ho and Heinke}(2009)}]{Ho09}
\bibinfo{author}{\bibfnamefont{W.~C.~G.} \bibnamefont{Ho}} \bibnamefont{and}
  \bibinfo{author}{\bibfnamefont{C.~O.} \bibnamefont{Heinke}},
  \bibinfo{journal}{Nature} \textbf{\bibinfo{volume}{462}}, \bibinfo{pages}{71}
  (\bibinfo{year}{2009}), \urlprefix\url{http://dx.doi.org/0.1038/nature08525}.

\bibitem[{\citenamefont{{Kumar} et~al.}(2012)\citenamefont{{Kumar},
  {Safi-Harb}, and {Gonzalez}}}]{Kumar2012}
\bibinfo{author}{\bibfnamefont{H.~S.} \bibnamefont{{Kumar}}},
  \bibinfo{author}{\bibfnamefont{S.}~\bibnamefont{{Safi-Harb}}},
  \bibnamefont{and} \bibinfo{author}{\bibfnamefont{M.~E.}
  \bibnamefont{{Gonzalez}}}, \bibinfo{journal}{\apj}
  \textbf{\bibinfo{volume}{754}}, \bibinfo{eid}{96} (\bibinfo{year}{2012}),
  \urlprefix\url{https://dx.doi.org/10.1088/0004-637X/754/2/96}.

\bibitem[{\citenamefont{{Safi-Harb} and {Kumar}}(2008)}]{SafiHarb08}
\bibinfo{author}{\bibfnamefont{S.}~\bibnamefont{{Safi-Harb}}} \bibnamefont{and}
  \bibinfo{author}{\bibfnamefont{H.~S.} \bibnamefont{{Kumar}}},
  \bibinfo{journal}{\apj} \textbf{\bibinfo{volume}{684}},
  \bibinfo{eid}{532-541} (\bibinfo{year}{2008}),
  \urlprefix\url{http://dx.doi.org/10.1086/590359}.

\bibitem[{\citenamefont{Zavlin et~al.}(1999)\citenamefont{Zavlin, Tr\"{u}mper,
  and Pavlov}}]{Zavlin:1999}
\bibinfo{author}{\bibfnamefont{V.~E.} \bibnamefont{Zavlin}},
  \bibinfo{author}{\bibfnamefont{J.}~\bibnamefont{Tr\"{u}mper}},
  \bibnamefont{and} \bibinfo{author}{\bibfnamefont{G.~G.}
  \bibnamefont{Pavlov}}, \bibinfo{journal}{Astrophys. J.}
  \textbf{\bibinfo{volume}{525}}, \bibinfo{pages}{959} (\bibinfo{year}{1999}),
  \urlprefix\url{http://dx.doi.org/10.1086/307919}.

\bibitem[{\citenamefont{Zavlin et~al.}(2000)\citenamefont{Zavlin, Pavlov,
  Sanwal, and Trümper}}]{Zavlin00}
\bibinfo{author}{\bibfnamefont{V.~E.} \bibnamefont{Zavlin}},
  \bibinfo{author}{\bibfnamefont{G.~G.} \bibnamefont{Pavlov}},
  \bibinfo{author}{\bibfnamefont{D.}~\bibnamefont{Sanwal}}, \bibnamefont{and}
  \bibinfo{author}{\bibfnamefont{J.}~\bibnamefont{Trümper}},
  \bibinfo{journal}{Astrophys. J. Lett.} \textbf{\bibinfo{volume}{540}},
  \bibinfo{pages}{L25} (\bibinfo{year}{2000}),
  \urlprefix\url{http://stacks.iop.org/1538-4357/540/i=1/a=L25}.

\bibitem[{\citenamefont{{Roger} et~al.}(1988)\citenamefont{{Roger}, {Milne},
  {Kesteven}, {Wellington}, and {Haynes}}}]{Roger98}
\bibinfo{author}{\bibfnamefont{R.~S.} \bibnamefont{{Roger}}},
  \bibinfo{author}{\bibfnamefont{D.~K.} \bibnamefont{{Milne}}},
  \bibinfo{author}{\bibfnamefont{M.~J.} \bibnamefont{{Kesteven}}},
  \bibinfo{author}{\bibfnamefont{K.~J.} \bibnamefont{{Wellington}}},
  \bibnamefont{and} \bibinfo{author}{\bibfnamefont{R.~F.}
  \bibnamefont{{Haynes}}}, \bibinfo{journal}{Astrophys. J}
  \textbf{\bibinfo{volume}{332}}, \bibinfo{pages}{940} (\bibinfo{year}{1988}),
  \urlprefix\url{http://dx.doi.org/10.1086/166703}.

\bibitem[{\citenamefont{Pavlov et~al.}(2002)\citenamefont{Pavlov, Zavlin,
  Sanwal, and Trumper}}]{Pavlov:2002b}
\bibinfo{author}{\bibfnamefont{G.~G.} \bibnamefont{Pavlov}},
  \bibinfo{author}{\bibfnamefont{V.~E.} \bibnamefont{Zavlin}},
  \bibinfo{author}{\bibfnamefont{D.}~\bibnamefont{Sanwal}}, \bibnamefont{and}
  \bibinfo{author}{\bibfnamefont{J.}~\bibnamefont{Trumper}},
  \bibinfo{journal}{Astrophys. J. Lett.} \textbf{\bibinfo{volume}{569}},
  \bibinfo{pages}{95} (\bibinfo{year}{2002}),
  \urlprefix\url{http://dx.doi.org/10.1086/340640}.

\bibitem[{\citenamefont{{Mereghetti} et~al.}(1996)\citenamefont{{Mereghetti},
  {Bignami}, and {Caraveo}}}]{Mereghetti96}
\bibinfo{author}{\bibfnamefont{S.}~\bibnamefont{{Mereghetti}}},
  \bibinfo{author}{\bibfnamefont{G.~F.} \bibnamefont{{Bignami}}},
  \bibnamefont{and} \bibinfo{author}{\bibfnamefont{P.~A.}
  \bibnamefont{{Caraveo}}}, \bibinfo{journal}{\apj}
  \textbf{\bibinfo{volume}{464}}, \bibinfo{pages}{842} (\bibinfo{year}{1996}),
  \urlprefix\url{http://dx.doi.org/10.1086/177370}.

\bibitem[{\citenamefont{{Zavlin} et~al.}(1998)\citenamefont{{Zavlin}, {Pavlov},
  and {Trumper}}}]{Zavlin98}
\bibinfo{author}{\bibfnamefont{V.~E.} \bibnamefont{{Zavlin}}},
  \bibinfo{author}{\bibfnamefont{G.~G.} \bibnamefont{{Pavlov}}},
  \bibnamefont{and}
  \bibinfo{author}{\bibfnamefont{J.}~\bibnamefont{{Trumper}}},
  \bibinfo{journal}{Astron. \& Astrophys.} \textbf{\bibinfo{volume}{331}},
  \bibinfo{pages}{821} (\bibinfo{year}{1998}),
  \urlprefix\url{http://arxiv.org/abs/astro-ph/9709267}.

\bibitem[{\citenamefont{Zavlin}(2007{\natexlab{a}})}]{Zavlin:2007nx}
\bibinfo{author}{\bibfnamefont{V.~E.} \bibnamefont{Zavlin}},
  \bibinfo{journal}{Astrophys. J. Lett.} \textbf{\bibinfo{volume}{665}},
  \bibinfo{pages}{143} (\bibinfo{year}{2007}{\natexlab{a}}),
  \urlprefix\url{http://dx.doi.org/10.1086/521300}.

\bibitem[{\citenamefont{Pavlov et~al.}(2004)\citenamefont{Pavlov, Sanwal, and
  Teter}}]{Pavlov04}
\bibinfo{author}{\bibfnamefont{G.~G.} \bibnamefont{Pavlov}},
  \bibinfo{author}{\bibfnamefont{D.}~\bibnamefont{Sanwal}}, \bibnamefont{and}
  \bibinfo{author}{\bibfnamefont{M.~A.} \bibnamefont{Teter}},
  \bibinfo{journal}{IAU Symp.} \textbf{\bibinfo{volume}{218}},
  \bibinfo{pages}{239} (\bibinfo{year}{2004}),
  \urlprefix\url{http://www.arxiv.org/abs/astro-ph/0311526}.

\bibitem[{\citenamefont{Tsuruta et~al.}(2009)\citenamefont{Tsuruta, Sadino,
  Kobelski, Teter, Liebmann, Takatsuka, Nomoto, and Umeda}}]{Tsuruta09}
\bibinfo{author}{\bibfnamefont{S.}~\bibnamefont{Tsuruta}},
  \bibinfo{author}{\bibfnamefont{J.}~\bibnamefont{Sadino}},
  \bibinfo{author}{\bibfnamefont{A.}~\bibnamefont{Kobelski}},
  \bibinfo{author}{\bibfnamefont{M.~A.} \bibnamefont{Teter}},
  \bibinfo{author}{\bibfnamefont{A.~C.} \bibnamefont{Liebmann}},
  \bibinfo{author}{\bibfnamefont{T.}~\bibnamefont{Takatsuka}},
  \bibinfo{author}{\bibfnamefont{K.}~\bibnamefont{Nomoto}}, \bibnamefont{and}
  \bibinfo{author}{\bibfnamefont{H.}~\bibnamefont{Umeda}},
  \bibinfo{journal}{Astrophys. J.} \textbf{\bibinfo{volume}{691}},
  \bibinfo{pages}{621} (\bibinfo{year}{2009}),
  \urlprefix\url{http://dx.doi.org/10.1088/0004-637X/691/1/621}.

\bibitem[{\citenamefont{Pavlov et~al.}(2001)\citenamefont{Pavlov, Zavlin,
  Sanwal, Burwitz, and Garmire}}]{Pavlov:2001hp}
\bibinfo{author}{\bibfnamefont{G.~G.} \bibnamefont{Pavlov}},
  \bibinfo{author}{\bibfnamefont{V.~E.} \bibnamefont{Zavlin}},
  \bibinfo{author}{\bibfnamefont{D.}~\bibnamefont{Sanwal}},
  \bibinfo{author}{\bibfnamefont{V.}~\bibnamefont{Burwitz}}, \bibnamefont{and}
  \bibinfo{author}{\bibfnamefont{G.}~\bibnamefont{Garmire}},
  \bibinfo{journal}{Astrophys. J. Lett.} \textbf{\bibinfo{volume}{552}},
  \bibinfo{pages}{129} (\bibinfo{year}{2001}),
  \urlprefix\url{http://dx.doi.org/10.1086/320342}.

\bibitem[{\citenamefont{{Gotthelf} et~al.}(2002)\citenamefont{{Gotthelf},
  {Halpern}, and {Dodson}}}]{Gotthelf02}
\bibinfo{author}{\bibfnamefont{E.~V.} \bibnamefont{{Gotthelf}}},
  \bibinfo{author}{\bibfnamefont{J.~P.} \bibnamefont{{Halpern}}},
  \bibnamefont{and} \bibinfo{author}{\bibfnamefont{R.}~\bibnamefont{{Dodson}}},
  \bibinfo{journal}{Astrophys. J. Lett.} \textbf{\bibinfo{volume}{567}},
  \bibinfo{pages}{125} (\bibinfo{year}{2002}),
  \urlprefix\url{http://dx.doi.org/10.1086/340109}.

\bibitem[{\citenamefont{McGowan et~al.}(2004)\citenamefont{McGowan, Zane,
  Cropper, Kennea, Cordova, Ho, Sasseen, and Vestrand}}]{McGowan04}
\bibinfo{author}{\bibfnamefont{K.~E.} \bibnamefont{McGowan}},
  \bibinfo{author}{\bibfnamefont{S.}~\bibnamefont{Zane}},
  \bibinfo{author}{\bibfnamefont{M.}~\bibnamefont{Cropper}},
  \bibinfo{author}{\bibfnamefont{J.~A.} \bibnamefont{Kennea}},
  \bibinfo{author}{\bibfnamefont{F.~A.} \bibnamefont{Cordova}},
  \bibinfo{author}{\bibfnamefont{C.}~\bibnamefont{Ho}},
  \bibinfo{author}{\bibfnamefont{T.}~\bibnamefont{Sasseen}}, \bibnamefont{and}
  \bibinfo{author}{\bibfnamefont{W.~T.} \bibnamefont{Vestrand}},
  \bibinfo{journal}{Astrophys. J.} \textbf{\bibinfo{volume}{600}},
  \bibinfo{pages}{343} (\bibinfo{year}{2004}),
  \urlprefix\url{http://dx.doi.org/10.1086/379787}.

\bibitem[{\citenamefont{Tian et~al.}(2008)\citenamefont{Tian, Leahy, Haverkorn,
  and Jiang}}]{Tian:2008tr}
\bibinfo{author}{\bibfnamefont{W.~W.} \bibnamefont{Tian}},
  \bibinfo{author}{\bibfnamefont{D.~A.} \bibnamefont{Leahy}},
  \bibinfo{author}{\bibfnamefont{M.}~\bibnamefont{Haverkorn}},
  \bibnamefont{and} \bibinfo{author}{\bibfnamefont{B.}~\bibnamefont{Jiang}},
  \bibinfo{journal}{Astrophys. J.} \textbf{\bibinfo{volume}{679}},
  \bibinfo{pages}{L85} (\bibinfo{year}{2008}),
  \urlprefix\url{http://dx.doi.org/10.1086/589506}.

\bibitem[{\citenamefont{Klochkov et~al.}(2015)\citenamefont{Klochkov,
  Suleimanov, Pühlhofer, Yakovlev, Santangelo, and Werner}}]{Klochkov:2014ola}
\bibinfo{author}{\bibfnamefont{D.}~\bibnamefont{Klochkov}},
  \bibinfo{author}{\bibfnamefont{V.}~\bibnamefont{Suleimanov}},
  \bibinfo{author}{\bibfnamefont{G.}~\bibnamefont{Pühlhofer}},
  \bibinfo{author}{\bibfnamefont{D.~G.} \bibnamefont{Yakovlev}},
  \bibinfo{author}{\bibfnamefont{A.}~\bibnamefont{Santangelo}},
  \bibnamefont{and} \bibinfo{author}{\bibfnamefont{K.}~\bibnamefont{Werner}},
  \bibinfo{journal}{Astron. Astrophys.} \textbf{\bibinfo{volume}{573}},
  \bibinfo{pages}{A53} (\bibinfo{year}{2015}),
  \urlprefix\url{http://dx.doi.org/10.1051/0004-6361/201424683}.

\bibitem[{\citenamefont{Kramer et~al.}(2003)\citenamefont{Kramer, Lyne, Hobbs,
  Lohmer, Carr, Jordan, and Wolszczan}}]{Kramer:2003au}
\bibinfo{author}{\bibfnamefont{M.}~\bibnamefont{Kramer}},
  \bibinfo{author}{\bibfnamefont{A.~G.} \bibnamefont{Lyne}},
  \bibinfo{author}{\bibfnamefont{G.}~\bibnamefont{Hobbs}},
  \bibinfo{author}{\bibfnamefont{O.}~\bibnamefont{Lohmer}},
  \bibinfo{author}{\bibfnamefont{P.}~\bibnamefont{Carr}},
  \bibinfo{author}{\bibfnamefont{C.}~\bibnamefont{Jordan}}, \bibnamefont{and}
  \bibinfo{author}{\bibfnamefont{A.}~\bibnamefont{Wolszczan}},
  \bibinfo{journal}{Astrophys. J. Lett.} \textbf{\bibinfo{volume}{593}},
  \bibinfo{pages}{31} (\bibinfo{year}{2003}),
  \urlprefix\url{http://dx.doi.org10.1086/378082}.

\bibitem[{\citenamefont{Zavlin and Pavlov}(2004)}]{Zavlin04c}
\bibinfo{author}{\bibfnamefont{V.~E.} \bibnamefont{Zavlin}} \bibnamefont{and}
  \bibinfo{author}{\bibfnamefont{G.~G.} \bibnamefont{Pavlov}},
  \bibinfo{journal}{Mem. Soc. Ast. It.} \textbf{\bibinfo{volume}{75}},
  \bibinfo{pages}{458} (\bibinfo{year}{2004}),
  \urlprefix\url{http://sait.oat.ts.astro.it/MmSAI/75/PDF/458.pdf}.

\bibitem[{\citenamefont{{McGowan} et~al.}(2003)\citenamefont{{McGowan},
  {Kennea}, {Zane}, {C{\'o}rdova}, {Cropper}, {Ho}, {Sasseen}, and
  {Vestrand}}}]{McGowan03}
\bibinfo{author}{\bibfnamefont{K.~E.} \bibnamefont{{McGowan}}},
  \bibinfo{author}{\bibfnamefont{J.~A.} \bibnamefont{{Kennea}}},
  \bibinfo{author}{\bibfnamefont{S.}~\bibnamefont{{Zane}}},
  \bibinfo{author}{\bibfnamefont{F.~A.} \bibnamefont{{C{\'o}rdova}}},
  \bibinfo{author}{\bibfnamefont{M.}~\bibnamefont{{Cropper}}},
  \bibinfo{author}{\bibfnamefont{C.}~\bibnamefont{{Ho}}},
  \bibinfo{author}{\bibfnamefont{T.}~\bibnamefont{{Sasseen}}},
  \bibnamefont{and} \bibinfo{author}{\bibfnamefont{W.~T.}
  \bibnamefont{{Vestrand}}}, \bibinfo{journal}{Astrophys. J.}
  \textbf{\bibinfo{volume}{591}}, \bibinfo{pages}{380} (\bibinfo{year}{2003}),
  \urlprefix\url{http://dx.doi.org/10.1086/375332}.

\bibitem[{\citenamefont{McGowan et~al.}(2006)\citenamefont{McGowan, Zane,
  Cropper, Vestrand, and Ho}}]{McGowan06}
\bibinfo{author}{\bibfnamefont{K.~E.} \bibnamefont{McGowan}},
  \bibinfo{author}{\bibfnamefont{S.}~\bibnamefont{Zane}},
  \bibinfo{author}{\bibfnamefont{M.}~\bibnamefont{Cropper}},
  \bibinfo{author}{\bibfnamefont{W.~T.} \bibnamefont{Vestrand}},
  \bibnamefont{and} \bibinfo{author}{\bibfnamefont{C.}~\bibnamefont{Ho}},
  \bibinfo{journal}{Astrophys. J.} \textbf{\bibinfo{volume}{639}},
  \bibinfo{pages}{377} (\bibinfo{year}{2006}),
  \urlprefix\url{http://dx.doi.org/10.1086/497327}.

\bibitem[{\citenamefont{Zavlin}(2007{\natexlab{b}})}]{Zavlin:2007nc}
\bibinfo{author}{\bibfnamefont{V.~E.} \bibnamefont{Zavlin}}, in
  \emph{\bibinfo{booktitle}{Neutron Stars and Pulsars: About 40 Years After the
  Discovery}} (\bibinfo{year}{2007}{\natexlab{b}}), vol. \bibinfo{volume}{357},
  p. \bibinfo{pages}{181},
  \urlprefix\url{http://dx.doi.org/10.1007/978-3-540-76965-1_9}.

\bibitem[{\citenamefont{Mignani et~al.}(2015)\citenamefont{Mignani, Moran,
  Shearer, Testa, Sowikowska, Rudak, Krzeszowki, and Kanbach}}]{Mignani15}
\bibinfo{author}{\bibfnamefont{R.~P.} \bibnamefont{Mignani}},
  \bibinfo{author}{\bibfnamefont{P.}~\bibnamefont{Moran}},
  \bibinfo{author}{\bibfnamefont{A.}~\bibnamefont{Shearer}},
  \bibinfo{author}{\bibfnamefont{V.}~\bibnamefont{Testa}},
  \bibinfo{author}{\bibfnamefont{A.}~\bibnamefont{Sowikowska}},
  \bibinfo{author}{\bibfnamefont{B.}~\bibnamefont{Rudak}},
  \bibinfo{author}{\bibfnamefont{K.}~\bibnamefont{Krzeszowki}},
  \bibnamefont{and} \bibinfo{author}{\bibfnamefont{G.}~\bibnamefont{Kanbach}},
  \bibinfo{journal}{Astron. Astrophys.} \textbf{\bibinfo{volume}{583}},
  \bibinfo{pages}{A105} (\bibinfo{year}{2015}), \eprint{1510.01057},
  \urlprefix\url{http://dx.doi.org/10.1051/0004-6361/201527082}.

\bibitem[{\citenamefont{Possenti et~al.}(1996)\citenamefont{Possenti,
  Mereghetti, and Colpi}}]{Possenti:1996em}
\bibinfo{author}{\bibfnamefont{A.}~\bibnamefont{Possenti}},
  \bibinfo{author}{\bibfnamefont{S.}~\bibnamefont{Mereghetti}},
  \bibnamefont{and} \bibinfo{author}{\bibfnamefont{M.}~\bibnamefont{Colpi}},
  \bibinfo{journal}{Astron. Astrophys.} \textbf{\bibinfo{volume}{313}},
  \bibinfo{pages}{565} (\bibinfo{year}{1996}),
  \urlprefix\url{http://adsabs.harvard.edu/abs/1996A%26A...313..565P}.

\bibitem[{\citenamefont{McLaughlin et~al.}(2002)\citenamefont{McLaughlin,
  Arzoumanian, Cordes, Backer, Lommen, Lorimer, and Zepka}}]{McLaughlin:2001kh}
\bibinfo{author}{\bibfnamefont{M.~A.} \bibnamefont{McLaughlin}},
  \bibinfo{author}{\bibfnamefont{Z.}~\bibnamefont{Arzoumanian}},
  \bibinfo{author}{\bibfnamefont{J.~M.} \bibnamefont{Cordes}},
  \bibinfo{author}{\bibfnamefont{D.~C.} \bibnamefont{Backer}},
  \bibinfo{author}{\bibfnamefont{A.~N.} \bibnamefont{Lommen}},
  \bibinfo{author}{\bibfnamefont{D.~R.} \bibnamefont{Lorimer}},
  \bibnamefont{and} \bibinfo{author}{\bibfnamefont{A.~F.} \bibnamefont{Zepka}},
  \bibinfo{journal}{Astrophys. J.} \textbf{\bibinfo{volume}{564}},
  \bibinfo{pages}{333} (\bibinfo{year}{2002}),
  \urlprefix\url{https://dx.doi.org/10.1086/324151}.

\bibitem[{\citenamefont{Kargaltsev et~al.}(2012)\citenamefont{Kargaltsev,
  Durant, Misanovic, and Pavlov}}]{Kargaltsev:2012yi}
\bibinfo{author}{\bibfnamefont{O.}~\bibnamefont{Kargaltsev}},
  \bibinfo{author}{\bibfnamefont{M.}~\bibnamefont{Durant}},
  \bibinfo{author}{\bibfnamefont{Z.}~\bibnamefont{Misanovic}},
  \bibnamefont{and} \bibinfo{author}{\bibfnamefont{G.}~\bibnamefont{Pavlov}},
  \bibinfo{journal}{Science} \textbf{\bibinfo{volume}{337}},
  \bibinfo{pages}{946} (\bibinfo{year}{2012}),
  \urlprefix\url{https://dx.doi.org/10.1126/science.1221378}.

\bibitem[{\citenamefont{{Vigan{\`o}} et~al.}(2013)\citenamefont{{Vigan{\`o}},
  {Rea}, {Pons}, {Perna}, {Aguilera}, and {Miralles}}}]{Vigano13}
\bibinfo{author}{\bibfnamefont{D.}~\bibnamefont{{Vigan{\`o}}}},
  \bibinfo{author}{\bibfnamefont{N.}~\bibnamefont{{Rea}}},
  \bibinfo{author}{\bibfnamefont{J.~A.} \bibnamefont{{Pons}}},
  \bibinfo{author}{\bibfnamefont{R.}~\bibnamefont{{Perna}}},
  \bibinfo{author}{\bibfnamefont{D.~N.} \bibnamefont{{Aguilera}}},
  \bibnamefont{and} \bibinfo{author}{\bibfnamefont{J.~A.}
  \bibnamefont{{Miralles}}}, \bibinfo{journal}{Mon. Not. R. Astron. Soc.}
  \textbf{\bibinfo{volume}{434}}, \bibinfo{pages}{123} (\bibinfo{year}{2013}),
  \urlprefix\url{https://dx.doi.org/10.1093/mnras/stt1008}.

\bibitem[{\citenamefont{Halpern and Wang}(1997)}]{Halpern1997a}
\bibinfo{author}{\bibfnamefont{J.~P.} \bibnamefont{Halpern}} \bibnamefont{and}
  \bibinfo{author}{\bibfnamefont{F.~Y.-H.} \bibnamefont{Wang}},
  \bibinfo{journal}{Astrophys. J.} \textbf{\bibinfo{volume}{477}},
  \bibinfo{pages}{905} (\bibinfo{year}{1997}),
  \urlprefix\url{http://iopscience.iop.org/article/10.1086/303743/pdf}.

\bibitem[{\citenamefont{Ho}(2007)}]{Ho:2007gs}
\bibinfo{author}{\bibfnamefont{W.~C.~G.} \bibnamefont{Ho}},
  \bibinfo{journal}{Mon. Not. Roy. Astron. Soc.}
  \textbf{\bibinfo{volume}{380}}, \bibinfo{pages}{71} (\bibinfo{year}{2007}),
  \urlprefix\url{http://dx.doi.org/10.1111/j.1365-2966.2007.12043.x}.

\bibitem[{\citenamefont{Pons et~al.}(2002)\citenamefont{Pons, Walter, Lattimer,
  Prakash, Neuhauser, and An}}]{Pons02}
\bibinfo{author}{\bibfnamefont{J.~A.} \bibnamefont{Pons}},
  \bibinfo{author}{\bibfnamefont{F.~M.} \bibnamefont{Walter}},
  \bibinfo{author}{\bibfnamefont{J.~M.} \bibnamefont{Lattimer}},
  \bibinfo{author}{\bibfnamefont{M.}~\bibnamefont{Prakash}},
  \bibinfo{author}{\bibfnamefont{R.}~\bibnamefont{Neuhauser}},
  \bibnamefont{and} \bibinfo{author}{\bibfnamefont{P.-h.} \bibnamefont{An}},
  \bibinfo{journal}{Astrophys. J.} \textbf{\bibinfo{volume}{564}},
  \bibinfo{pages}{981} (\bibinfo{year}{2002}),
  \urlprefix\url{http://dx.doi.org/10.1086/324296}.

\bibitem[{\citenamefont{Burwitz et~al.}(2003)\citenamefont{Burwitz, Haberl,
  Neuhaeuser, Predehl, Truemper, and Zavlin}}]{Burwitz03}
\bibinfo{author}{\bibfnamefont{V.}~\bibnamefont{Burwitz}},
  \bibinfo{author}{\bibfnamefont{F.}~\bibnamefont{Haberl}},
  \bibinfo{author}{\bibfnamefont{R.}~\bibnamefont{Neuhaeuser}},
  \bibinfo{author}{\bibfnamefont{P.}~\bibnamefont{Predehl}},
  \bibinfo{author}{\bibfnamefont{J.}~\bibnamefont{Truemper}}, \bibnamefont{and}
  \bibinfo{author}{\bibfnamefont{V.~E.} \bibnamefont{Zavlin}},
  \bibinfo{journal}{Astron. Astrophys.} \textbf{\bibinfo{volume}{399}},
  \bibinfo{pages}{1109} (\bibinfo{year}{2003}),
  \urlprefix\url{http://dx.doi.org/10.1051/0004-6361:20021747}.

\bibitem[{\citenamefont{Pavlov and Zavlin}(2003)}]{Pavlov:2003da}
\bibinfo{author}{\bibfnamefont{G.~G.} \bibnamefont{Pavlov}} \bibnamefont{and}
  \bibinfo{author}{\bibfnamefont{V.~E.} \bibnamefont{Zavlin}}, in
  \emph{\bibinfo{booktitle}{{Proceedings 21st Texas Symposium on Relativistic
  Astrophysics. Edited by R. Bandiera, R. Maiolino and F. Mannucci. Singapore,
  World Scientific, 2003. p. 319}}} (\bibinfo{year}{2003}),
  \urlprefix\url{http://www.arxiv.org/abs/astro-ph/0305435}.

\bibitem[{\citenamefont{de~Vries et~al.}(2004)\citenamefont{de~Vries, Vink,
  Mendez, and Verbunt}}]{deVries:2004jf}
\bibinfo{author}{\bibfnamefont{C.~P.} \bibnamefont{de~Vries}},
  \bibinfo{author}{\bibfnamefont{J.}~\bibnamefont{Vink}},
  \bibinfo{author}{\bibfnamefont{M.}~\bibnamefont{Mendez}}, \bibnamefont{and}
  \bibinfo{author}{\bibfnamefont{F.}~\bibnamefont{Verbunt}},
  \bibinfo{journal}{Astron. Astrophys.} \textbf{\bibinfo{volume}{415}},
  \bibinfo{pages}{L31} (\bibinfo{year}{2004}),
  \urlprefix\url{http://dx.doi.org/10.1051/0004-6361:20040009}.

\bibitem[{\citenamefont{{Kaplan} et~al.}(2002)\citenamefont{{Kaplan},
  {Kulkarni}, {van Kerkwijk}, and {Marshall}}}]{Kaplan02b}
\bibinfo{author}{\bibfnamefont{D.~L.} \bibnamefont{{Kaplan}}},
  \bibinfo{author}{\bibfnamefont{S.~R.} \bibnamefont{{Kulkarni}}},
  \bibinfo{author}{\bibfnamefont{M.~H.} \bibnamefont{{van Kerkwijk}}},
  \bibnamefont{and} \bibinfo{author}{\bibfnamefont{H.~L.}
  \bibnamefont{{Marshall}}}, \bibinfo{journal}{Astrophys. J. Lett.}
  \textbf{\bibinfo{volume}{570}}, \bibinfo{pages}{79} (\bibinfo{year}{2002}),
  \urlprefix\url{http://dx.doi.org/10.1086/341102}.

\bibitem[{\citenamefont{Kaplan et~al.}(2003)\citenamefont{Kaplan, van Kerkwijk,
  Marshall, Jacoby, Kulkarni, and Frail}}]{Kaplan:2003hj}
\bibinfo{author}{\bibfnamefont{D.~L.} \bibnamefont{Kaplan}},
  \bibinfo{author}{\bibfnamefont{M.~H.} \bibnamefont{van Kerkwijk}},
  \bibinfo{author}{\bibfnamefont{H.~L.} \bibnamefont{Marshall}},
  \bibinfo{author}{\bibfnamefont{B.~A.} \bibnamefont{Jacoby}},
  \bibinfo{author}{\bibfnamefont{S.~R.} \bibnamefont{Kulkarni}},
  \bibnamefont{and} \bibinfo{author}{\bibfnamefont{D.~A.} \bibnamefont{Frail}},
  \bibinfo{journal}{Astrophys. J.} \textbf{\bibinfo{volume}{590}},
  \bibinfo{pages}{1008} (\bibinfo{year}{2003}),
  \urlprefix\url{http://dx.doi.org/10.1086/375052}.

\bibitem[{\citenamefont{Lim et~al.}(2017)\citenamefont{Lim, Hyun, and
  Lee}}]{Lim15}
\bibinfo{author}{\bibfnamefont{Y.}~\bibnamefont{Lim}},
  \bibinfo{author}{\bibfnamefont{C.~H.} \bibnamefont{Hyun}}, \bibnamefont{and}
  \bibinfo{author}{\bibfnamefont{C.-H.} \bibnamefont{Lee}},
  \bibinfo{journal}{Int. J. Mod. Phys. E} \textbf{\bibinfo{volume}{26}},
  \bibinfo{pages}{1750015} (\bibinfo{year}{2017}),
  \urlprefix\url{https://dx.doi.org/10.1142/S021830131750015X}.

\bibitem[{\citenamefont{Esposito et~al.}(2008)\citenamefont{Esposito, De~Luca,
  Tiengo, Paizis, Mereghetti, and Caraveo}}]{Esposito08}
\bibinfo{author}{\bibfnamefont{P.}~\bibnamefont{Esposito}},
  \bibinfo{author}{\bibfnamefont{A.}~\bibnamefont{De~Luca}},
  \bibinfo{author}{\bibfnamefont{A.}~\bibnamefont{Tiengo}},
  \bibinfo{author}{\bibfnamefont{A.}~\bibnamefont{Paizis}},
  \bibinfo{author}{\bibfnamefont{S.}~\bibnamefont{Mereghetti}},
  \bibnamefont{and} \bibinfo{author}{\bibfnamefont{P.~A.}
  \bibnamefont{Caraveo}}, \bibinfo{journal}{Mon. Not. Roy. Astron. Soc.}
  \textbf{\bibinfo{volume}{384}}, \bibinfo{pages}{225} (\bibinfo{year}{2008}),
  \urlprefix\url{http://dx.doi.org/10.1111/j.1365-2966.2007.12677.x}.

\bibitem[{\citenamefont{Zavlin et~al.}(2004)\citenamefont{Zavlin, Pavlov, and
  Sanwal}}]{Zavlin04a}
\bibinfo{author}{\bibfnamefont{V.~E.} \bibnamefont{Zavlin}},
  \bibinfo{author}{\bibfnamefont{G.~G.} \bibnamefont{Pavlov}},
  \bibnamefont{and} \bibinfo{author}{\bibfnamefont{D.}~\bibnamefont{Sanwal}},
  \bibinfo{journal}{Astrophys. J.} \textbf{\bibinfo{volume}{606}},
  \bibinfo{pages}{444} (\bibinfo{year}{2004}),
  \urlprefix\url{http://dx.doi.org/10.1086/382725}.

\bibitem[{\citenamefont{Akmal et~al.}(1998)\citenamefont{Akmal, Pandharipande,
  and Ravenhall}}]{Akmal98eo}
\bibinfo{author}{\bibfnamefont{A.}~\bibnamefont{Akmal}},
  \bibinfo{author}{\bibfnamefont{V.~R.} \bibnamefont{Pandharipande}},
  \bibnamefont{and} \bibinfo{author}{\bibfnamefont{D.~G.}
  \bibnamefont{Ravenhall}}, \bibinfo{journal}{Phys. Rev. C}
  \textbf{\bibinfo{volume}{58}}, \bibinfo{pages}{1804} (\bibinfo{year}{1998}),
  \urlprefix\url{http://dx.doi.org/10.1103/PhysRevC.58.1804}.

\bibitem[{\citenamefont{Flowers et~al.}(1976)\citenamefont{Flowers, Ruderman,
  and Sutherland}}]{Flowers76a}
\bibinfo{author}{\bibfnamefont{E.}~\bibnamefont{Flowers}},
  \bibinfo{author}{\bibfnamefont{M.}~\bibnamefont{Ruderman}}, \bibnamefont{and}
  \bibinfo{author}{\bibfnamefont{P.}~\bibnamefont{Sutherland}},
  \bibinfo{journal}{Astrophys. J.} \textbf{\bibinfo{volume}{205}},
  \bibinfo{pages}{541} (\bibinfo{year}{1976}),
  \urlprefix\url{http://dx.doi.org/10.1086/154308}.

\bibitem[{\citenamefont{Leinson and Perez}(2006)}]{Leinson06ne}
\bibinfo{author}{\bibfnamefont{L.~B.} \bibnamefont{Leinson}} \bibnamefont{and}
  \bibinfo{author}{\bibfnamefont{A.}~\bibnamefont{Perez}},
  \bibinfo{journal}{arXiv:astro-ph/0606653}  (\bibinfo{year}{2006}),
  \urlprefix\url{http://arxiv.org/abs/astro-ph/0606653}.

\bibitem[{\citenamefont{{Leinson} and {P{\'e}rez}}(2006)}]{Leinson06vc}
\bibinfo{author}{\bibfnamefont{L.~B.} \bibnamefont{{Leinson}}}
  \bibnamefont{and}
  \bibinfo{author}{\bibfnamefont{A.}~\bibnamefont{{P{\'e}rez}}},
  \bibinfo{journal}{Phys. Lett. B} \textbf{\bibinfo{volume}{638}},
  \bibinfo{pages}{114} (\bibinfo{year}{2006}),
  \urlprefix\url{http://dx.doi.org/10.1016/j.physletb.2006.05.036}.

\bibitem[{\citenamefont{Steiner and {Reddy}}(2009)}]{Steiner09sr}
\bibinfo{author}{\bibfnamefont{A.~W.} \bibnamefont{Steiner}} \bibnamefont{and}
  \bibinfo{author}{\bibfnamefont{S.}~\bibnamefont{{Reddy}}},
  \bibinfo{journal}{Phys. Rev. C} \textbf{\bibinfo{volume}{79}},
  \bibinfo{pages}{015802} (\bibinfo{year}{2009}),
  \urlprefix\url{http://dx.doi.org/10.1103/PhysRevC.79.015802}.

\bibitem[{\citenamefont{Leinson}(2010)}]{Leinson:2009nu}
\bibinfo{author}{\bibfnamefont{L.~B.} \bibnamefont{Leinson}},
  \bibinfo{journal}{Phys. Rev. C} \textbf{\bibinfo{volume}{81}},
  \bibinfo{pages}{025501} (\bibinfo{year}{2010}),
  \urlprefix\url{https://dx.doi.org/10.1103/PhysRevC.81.025501}.

\bibitem[{\citenamefont{Isobe et~al.}(1990)\citenamefont{Isobe, Feigelson,
  Akritas, and Babu}}]{Isobe90}
\bibinfo{author}{\bibfnamefont{T.}~\bibnamefont{Isobe}},
  \bibinfo{author}{\bibfnamefont{E.~D.} \bibnamefont{Feigelson}},
  \bibinfo{author}{\bibfnamefont{M.~G.} \bibnamefont{Akritas}},
  \bibnamefont{and} \bibinfo{author}{\bibfnamefont{G.~J.} \bibnamefont{Babu}},
  \bibinfo{journal}{Astrophys. J.} \textbf{\bibinfo{volume}{364}},
  \bibinfo{pages}{104} (\bibinfo{year}{1990}),
  \urlprefix\url{http://dx.doi.org/10.1086/169390}.

\bibitem[{\citenamefont{Draper and Yang}(1997)}]{Draper97}
\bibinfo{author}{\bibfnamefont{N.~R.} \bibnamefont{Draper}} \bibnamefont{and}
  \bibinfo{author}{\bibfnamefont{Y.}~\bibnamefont{Yang}},
  \bibinfo{journal}{Comp. Stat. and Data Anal.} \textbf{\bibinfo{volume}{23}},
  \bibinfo{pages}{355} (\bibinfo{year}{1997}),
  \urlprefix\url{http://dx.doi.org/10.1016/S0167-9473(96)00037-0}.

\bibitem[{\citenamefont{Str\"{o}mberg}(1990)}]{Stromberg40}
\bibinfo{author}{\bibfnamefont{G.}~\bibnamefont{Str\"{o}mberg}},
  \bibinfo{journal}{Astrophys. J.} \textbf{\bibinfo{volume}{92}},
  \bibinfo{pages}{156} (\bibinfo{year}{1990}),
  \urlprefix\url{http://dx.doi.org/10.1086/144209}.

\bibitem[{\citenamefont{Steiner et~al.}(2010)\citenamefont{Steiner, Lattimer,
  and {Brown}}}]{Steiner10te}
\bibinfo{author}{\bibfnamefont{A.~W.} \bibnamefont{Steiner}},
  \bibinfo{author}{\bibfnamefont{J.~M.} \bibnamefont{Lattimer}},
  \bibnamefont{and} \bibinfo{author}{\bibfnamefont{E.~F.}
  \bibnamefont{{Brown}}}, \bibinfo{journal}{Astrophys. J.}
  \textbf{\bibinfo{volume}{722}}, \bibinfo{pages}{33} (\bibinfo{year}{2010}),
  \urlprefix\url{http://dx.doi.org/10.1088/0004-637X/722/1/33}.

\bibitem[{\citenamefont{{Page} et~al.}(2011)\citenamefont{{Page}, {Prakash},
  {Lattimer}, and {Steiner}}}]{Page11rc}
\bibinfo{author}{\bibfnamefont{D.}~\bibnamefont{{Page}}},
  \bibinfo{author}{\bibfnamefont{M.}~\bibnamefont{{Prakash}}},
  \bibinfo{author}{\bibfnamefont{J.~M.} \bibnamefont{{Lattimer}}},
  \bibnamefont{and} \bibinfo{author}{\bibfnamefont{A.~W.}
  \bibnamefont{{Steiner}}}, \bibinfo{journal}{Phys. Rev. Lett.}
  \textbf{\bibinfo{volume}{106}}, \bibinfo{pages}{081101}
  (\bibinfo{year}{2011}),
  \urlprefix\url{http://dx.doi.org/10.1103/PhysRevLett.106.081101}.

\bibitem[{\citenamefont{Shternin et~al.}(2011)\citenamefont{Shternin, Yakovlev,
  Heinke, Ho, and Patnaude}}]{Shternin11}
\bibinfo{author}{\bibfnamefont{P.~S.} \bibnamefont{Shternin}},
  \bibinfo{author}{\bibfnamefont{D.~G.} \bibnamefont{Yakovlev}},
  \bibinfo{author}{\bibfnamefont{C.~O.} \bibnamefont{Heinke}},
  \bibinfo{author}{\bibfnamefont{W.~C.~G.} \bibnamefont{Ho}}, \bibnamefont{and}
  \bibinfo{author}{\bibfnamefont{D.~J.} \bibnamefont{Patnaude}},
  \bibinfo{journal}{Mon. Not. R. Astron. Soc. Lett.}
  \textbf{\bibinfo{volume}{412}}, \bibinfo{pages}{108} (\bibinfo{year}{2011}),
  \urlprefix\url{http://dx.doi.org/10.1111/j.1745-3933.2011.01015.x}.

\bibitem[{\citenamefont{Blaschke et~al.}(2012)\citenamefont{Blaschke,
  Grigorian, Voskresensky, and Weber}}]{Blaschke12}
\bibinfo{author}{\bibfnamefont{D.}~\bibnamefont{Blaschke}},
  \bibinfo{author}{\bibfnamefont{H.}~\bibnamefont{Grigorian}},
  \bibinfo{author}{\bibfnamefont{D.~N.} \bibnamefont{Voskresensky}},
  \bibnamefont{and} \bibinfo{author}{\bibfnamefont{F.}~\bibnamefont{Weber}},
  \bibinfo{journal}{Phys. Rev. C} \textbf{\bibinfo{volume}{85}},
  \bibinfo{pages}{022802} (\bibinfo{year}{2012}),
  \urlprefix\url{http://dx.doi.org/10.1103/PhysRevC.85.022802}.

\bibitem[{\citenamefont{Blaschke et~al.}(2013)\citenamefont{Blaschke,
  Grigorian, and Voskresensky}}]{Blaschke13}
\bibinfo{author}{\bibfnamefont{D.}~\bibnamefont{Blaschke}},
  \bibinfo{author}{\bibfnamefont{H.}~\bibnamefont{Grigorian}},
  \bibnamefont{and} \bibinfo{author}{\bibfnamefont{D.~N.}
  \bibnamefont{Voskresensky}}, \bibinfo{journal}{Phys. Rev. C}
  \textbf{\bibinfo{volume}{88}}, \bibinfo{pages}{065805}
  (\bibinfo{year}{2013}),
  \urlprefix\url{http://dx.doi.org/10.1103/PhysRevC.88.065805}.

\bibitem[{\citenamefont{Ho et~al.}(2015)\citenamefont{Ho, Elshamouty, Heinke,
  and Potekhin}}]{Ho15}
\bibinfo{author}{\bibfnamefont{W.~C.~G.} \bibnamefont{Ho}},
  \bibinfo{author}{\bibfnamefont{K.~G.} \bibnamefont{Elshamouty}},
  \bibinfo{author}{\bibfnamefont{C.~O.} \bibnamefont{Heinke}},
  \bibnamefont{and} \bibinfo{author}{\bibfnamefont{A.~Y.}
  \bibnamefont{Potekhin}}, \bibinfo{journal}{Phys. Rev. C}
  \textbf{\bibinfo{volume}{91}}, \bibinfo{pages}{015806}
  (\bibinfo{year}{2015}),
  \urlprefix\url{http://dx.doi.org/10.1103/PhysRevC.91.015806}.

\bibitem[{\citenamefont{{Elshamouty} et~al.}(2013)\citenamefont{{Elshamouty},
  {Heinke}, {Sivakoff}, {Ho}, {Shternin}, {Yakovlev}, {Patnaude}, and
  {David}}}]{Elshamouty13}
\bibinfo{author}{\bibfnamefont{K.~G.} \bibnamefont{{Elshamouty}}},
  \bibinfo{author}{\bibfnamefont{C.~O.} \bibnamefont{{Heinke}}},
  \bibinfo{author}{\bibfnamefont{G.~R.} \bibnamefont{{Sivakoff}}},
  \bibinfo{author}{\bibfnamefont{W.~C.~G.} \bibnamefont{{Ho}}},
  \bibinfo{author}{\bibfnamefont{P.~S.} \bibnamefont{{Shternin}}},
  \bibinfo{author}{\bibfnamefont{D.~G.} \bibnamefont{{Yakovlev}}},
  \bibinfo{author}{\bibfnamefont{D.~J.} \bibnamefont{{Patnaude}}},
  \bibnamefont{and} \bibinfo{author}{\bibfnamefont{L.}~\bibnamefont{{David}}},
  \bibinfo{journal}{\apj} \textbf{\bibinfo{volume}{777}}, \bibinfo{eid}{22}
  (\bibinfo{year}{2013}),
  \urlprefix\url{http://dx.doi.org/10.1088/0004-637X/777/1/22}.

\bibitem[{\citenamefont{{Posselt} et~al.}(2013)\citenamefont{{Posselt},
  {Pavlov}, {Suleimanov}, and {Kargaltsev}}}]{Posselt13}
\bibinfo{author}{\bibfnamefont{B.}~\bibnamefont{{Posselt}}},
  \bibinfo{author}{\bibfnamefont{G.~G.} \bibnamefont{{Pavlov}}},
  \bibinfo{author}{\bibfnamefont{V.}~\bibnamefont{{Suleimanov}}},
  \bibnamefont{and}
  \bibinfo{author}{\bibfnamefont{O.}~\bibnamefont{{Kargaltsev}}},
  \bibinfo{journal}{\apj} \textbf{\bibinfo{volume}{779}}, \bibinfo{eid}{186}
  (\bibinfo{year}{2013}),
  \urlprefix\url{http://dx.doi.org/10.1088/0004-637X/779/2/186}.

\bibitem[{\citenamefont{Beznogov and
  Yakovlev}(2015{\natexlab{a}})}]{Beznogov15a}
\bibinfo{author}{\bibfnamefont{M.~V.} \bibnamefont{Beznogov}} \bibnamefont{and}
  \bibinfo{author}{\bibfnamefont{D.~G.} \bibnamefont{Yakovlev}},
  \bibinfo{journal}{Mon. Not. Roy. Astron. Soc.}
  \textbf{\bibinfo{volume}{447}}, \bibinfo{pages}{1598}
  (\bibinfo{year}{2015}{\natexlab{a}}),
  \urlprefix\url{http://dx.doi.org/10.1093/mnras/stu2506}.

\bibitem[{\citenamefont{Beznogov and
  Yakovlev}(2015{\natexlab{b}})}]{Beznogov15b}
\bibinfo{author}{\bibfnamefont{M.~V.} \bibnamefont{Beznogov}} \bibnamefont{and}
  \bibinfo{author}{\bibfnamefont{D.~G.} \bibnamefont{Yakovlev}},
  \bibinfo{journal}{Mon. Not. Roy. Astron. Soc.}
  \textbf{\bibinfo{volume}{452}}, \bibinfo{pages}{540}
  (\bibinfo{year}{2015}{\natexlab{b}}),
  \urlprefix\url{http://dx.doi.org/10.1093/mnras/stv1293}.

\bibitem[{\citenamefont{Heinke et~al.}(2007)\citenamefont{Heinke, Jonker,
  Wijnands, and Taam}}]{Heinke07}
\bibinfo{author}{\bibfnamefont{C.~O.} \bibnamefont{Heinke}},
  \bibinfo{author}{\bibfnamefont{P.~G.} \bibnamefont{Jonker}},
  \bibinfo{author}{\bibfnamefont{R.}~\bibnamefont{Wijnands}}, \bibnamefont{and}
  \bibinfo{author}{\bibfnamefont{R.~E.} \bibnamefont{Taam}},
  \bibinfo{journal}{Astrophys. J.} \textbf{\bibinfo{volume}{660}},
  \bibinfo{pages}{1424} (\bibinfo{year}{2007}),
  \urlprefix\url{http://dx.doi.org/10.1086/513140}.

\bibitem[{\citenamefont{Potekhin et~al.}(2015)\citenamefont{Potekhin, Pons, and
  Page}}]{Potekhin:2015qsa}
\bibinfo{author}{\bibfnamefont{A.~Y.} \bibnamefont{Potekhin}},
  \bibinfo{author}{\bibfnamefont{J.~A.} \bibnamefont{Pons}}, \bibnamefont{and}
  \bibinfo{author}{\bibfnamefont{D.}~\bibnamefont{Page}},
  \bibinfo{journal}{Space Sci. Rev.} \textbf{\bibinfo{volume}{191}},
  \bibinfo{pages}{239} (\bibinfo{year}{2015}),
  \urlprefix\url{https://dx.doi.org/10.1007/s11214-015-0180-9}.

\bibitem[{\citenamefont{de~Carvalho et~al.}(2015)\citenamefont{de~Carvalho,
  Negreiros, Orsaria, Contrera, Weber, and Spinella}}]{deCarvalho15}
\bibinfo{author}{\bibfnamefont{S.~M.} \bibnamefont{de~Carvalho}},
  \bibinfo{author}{\bibfnamefont{R.}~\bibnamefont{Negreiros}},
  \bibinfo{author}{\bibfnamefont{M.}~\bibnamefont{Orsaria}},
  \bibinfo{author}{\bibfnamefont{G.~A.} \bibnamefont{Contrera}},
  \bibinfo{author}{\bibfnamefont{F.}~\bibnamefont{Weber}}, \bibnamefont{and}
  \bibinfo{author}{\bibfnamefont{W.}~\bibnamefont{Spinella}},
  \bibinfo{journal}{Phys. Rev. C} \textbf{\bibinfo{volume}{92}},
  \bibinfo{pages}{035810} (\bibinfo{year}{2015}),
  \urlprefix\url{http://dx.doi.org/10.1103/PhysRevC.92.035810}.

\bibitem[{\citenamefont{Grigorian et~al.}(2015)\citenamefont{Grigorian,
  Blaschke, and Voskresensky}}]{Grigorian15}
\bibinfo{author}{\bibfnamefont{H.}~\bibnamefont{Grigorian}},
  \bibinfo{author}{\bibfnamefont{D.}~\bibnamefont{Blaschke}}, \bibnamefont{and}
  \bibinfo{author}{\bibfnamefont{D.~N.} \bibnamefont{Voskresensky}},
  \bibinfo{journal}{Phys. Part. Nucl.} \textbf{\bibinfo{volume}{46}},
  \bibinfo{pages}{849} (\bibinfo{year}{2015}),
  \urlprefix\url{http://dx.doi.org/10.1134/S1063779615050111}.

\bibitem[{\citenamefont{Grigorian et~al.}(2016)\citenamefont{Grigorian,
  Voskresensky, and Blaschke}}]{Grigorian16}
\bibinfo{author}{\bibfnamefont{H.}~\bibnamefont{Grigorian}},
  \bibinfo{author}{\bibfnamefont{D.~N.} \bibnamefont{Voskresensky}},
  \bibnamefont{and} \bibinfo{author}{\bibfnamefont{D.}~\bibnamefont{Blaschke}},
  \bibinfo{journal}{Eur. Phys. J. A} \textbf{\bibinfo{volume}{52}},
  \bibinfo{pages}{67} (\bibinfo{year}{2016}),
  \urlprefix\url{http://dx.doi.org/10.1140/epja/i2016-16067-4}.

\bibitem[{\citenamefont{Sedrakian}(2016{\natexlab{a}})}]{Sedrakian16}
\bibinfo{author}{\bibfnamefont{A.}~\bibnamefont{Sedrakian}},
  \bibinfo{journal}{Eur. Phys. J. A} \textbf{\bibinfo{volume}{52}},
  \bibinfo{pages}{44} (\bibinfo{year}{2016}{\natexlab{a}}),
  \urlprefix\url{http://dx.doi.org/10.1140/epja/i2016-16044-y}.

\bibitem[{\citenamefont{Sedrakian}(2016{\natexlab{b}})}]{Sedrakian16b}
\bibinfo{author}{\bibfnamefont{A.}~\bibnamefont{Sedrakian}},
  \bibinfo{journal}{Phys. Rev. D} \textbf{\bibinfo{volume}{93}},
  \bibinfo{pages}{065044} (\bibinfo{year}{2016}{\natexlab{b}}),
  \urlprefix\url{http://dx.doi.org/10.1103/PhysRevD.93.065044}.

\bibitem[{\citenamefont{Noda et~al.}(2013)\citenamefont{Noda, Hashimoto,
  Yasutake, Maruyama, Tatsumi, and Fujimoto}}]{Noda13}
\bibinfo{author}{\bibfnamefont{T.}~\bibnamefont{Noda}},
  \bibinfo{author}{\bibfnamefont{M.-A.} \bibnamefont{Hashimoto}},
  \bibinfo{author}{\bibfnamefont{N.}~\bibnamefont{Yasutake}},
  \bibinfo{author}{\bibfnamefont{T.}~\bibnamefont{Maruyama}},
  \bibinfo{author}{\bibfnamefont{T.}~\bibnamefont{Tatsumi}}, \bibnamefont{and}
  \bibinfo{author}{\bibfnamefont{M.}~\bibnamefont{Fujimoto}},
  \bibinfo{journal}{Astrophys. J.} \textbf{\bibinfo{volume}{765}},
  \bibinfo{pages}{1} (\bibinfo{year}{2013}),
  \urlprefix\url{http://dx.doi.org/10.1088/0004-637X/765/1/1}.

\bibitem[{\citenamefont{Ofengeim et~al.}(2015)\citenamefont{Ofengeim, Kaminker,
  Klochkov, Suleimanov, and Yakovlev}}]{Ofengeim15}
\bibinfo{author}{\bibfnamefont{D.~D.} \bibnamefont{Ofengeim}},
  \bibinfo{author}{\bibfnamefont{A.~D.} \bibnamefont{Kaminker}},
  \bibinfo{author}{\bibfnamefont{D.}~\bibnamefont{Klochkov}},
  \bibinfo{author}{\bibfnamefont{V.}~\bibnamefont{Suleimanov}},
  \bibnamefont{and} \bibinfo{author}{\bibfnamefont{D.~G.}
  \bibnamefont{Yakovlev}}, \bibinfo{journal}{Mon. Not. R. Astroc. Soc.}
  \textbf{\bibinfo{volume}{454}}, \bibinfo{pages}{2668} (\bibinfo{year}{2015}),
  \urlprefix\url{http://dx.doi.org/10.1093/mnras/stv2204}.

\bibitem[{\citenamefont{Suleimanov et~al.}(2017)\citenamefont{Suleimanov,
  Klochkov, Poutanen, and Werner}}]{Suleimanov:2017brq}
\bibinfo{author}{\bibfnamefont{V.~F.} \bibnamefont{Suleimanov}},
  \bibinfo{author}{\bibfnamefont{D.}~\bibnamefont{Klochkov}},
  \bibinfo{author}{\bibfnamefont{J.}~\bibnamefont{Poutanen}}, \bibnamefont{and}
  \bibinfo{author}{\bibfnamefont{K.}~\bibnamefont{Werner}},
  \bibinfo{journal}{Astron. Astrophys.} \textbf{\bibinfo{volume}{600}},
  \bibinfo{pages}{A43} (\bibinfo{year}{2017}),
  \urlprefix\url{https://dx.doi.org/10.1051/0004-6361/201630028}.

\end{thebibliography}

\end{document}